%% file: main.tex
\documentclass[fleqn,usenatbib]{mnras}
 
\usepackage[T1]{fontenc}
\usepackage{ae,aecompl}

\usepackage{graphicx}	
\usepackage{amsmath}	
\usepackage{amssymb}	
\usepackage{float}
\usepackage[dvipsnames,usenames,table]{xcolor}
\usepackage{array,multirow} 

\input{macros/newcommands}

\title[Constraining the TP-AGB phase in the SMC]{Constraining the thermally-pulsing asymptotic giant branch phase with resolved stellar populations in the Small Magellanic Cloud }
 
\author[G. Pastorelli et al.]{
Giada Pastorelli$^1$\thanks{E-mail: giada.pastorelli@unipd.it}, 
Paola Marigo$^1$, 
L\'eo Girardi$^2$, 
Yang Chen$^1$, 
Stefano Rubele$^{1,2}$, \newauthor 
Michele Trabucchi$^1$, 
Bernhard Aringer$^1$, 
Sara Bladh$^{1,3}$, 
Alessandro Bressan$^4$, \newauthor 
Josefina Montalb\'an$^1$, 
Martha L.\ Boyer$^5$, 
Julianne J.\ Dalcanton$^6$,
Kjell Eriksson$^3$, \newauthor 
Martin A.T.\ Groenewegen$^7$, 
Susanne H\"{o}fner$^3$,
Thomas Lebzelter$^{8}$, 
Ambra Nanni$^{1}$, \newauthor  
Philip Rosenfield$^{9}$,  
Peter R.\ Wood$^{10}$ and 
Maria-Rosa L.\ Cioni$^{11}$ 
\\
$^1$ Dipartimento di Fisica e Astronomia Galileo Galilei, Universit\`a di Padova, Vicolo dell'Osservatorio 3, I-35122 Padova, Italy \\
$^2$ Osservatorio Astronomico di Padova -- INAF, Vicolo dell'Osservatorio 5, I-35122 Padova, Italy \\
$^3$ Theoretical Astrophysics, Department of Physics and Astronomy, Uppsala
University, Box 516, SE-751 20 Uppsala, Sweden\\
$^4$ SISSA, via Bonomea 365, I-34136 Trieste, Italy \\
$^5$ STScI, 3700 San Martin Drive, Baltimore, MD 21218, USA\\ 
$^6$ Department of Astronomy, University of Washington, Box 351580, Seattle, WA 98195, USA \\
$^7$ Koninklijke Sterrenwacht van Belgi\"e, Ringlaan 3, 1180 Brussel, Belgium \\ 
$^8$ University of Vienna, Department of Astrophysics, Tuerkenschanzstrasse 17, A1180 Vienna, Austria \\
$^9$ Eureka Scientific, Inc., 2452 Delmer Street, Oakland CA 94602, USA\\
$^{10}$ Research School of Astronomy and Astrophysics, Australian National University, Canberra, ACT2611, Australia\\
$^{11}$ Leibniz-Instit\"{u}t f\"{u}r Astrophysik Potsdam, An der Sternwarte 16, D-14482 Potsdam, Germany  
}

\date{Accepted 2019 March 11. Received 2019 March 8; in original form 2018 November 28}

\pubyear{2019}
\begin{document}
\label{firstpage}
\pagerange{\pageref{firstpage}--\pageref{lastpage}}
\maketitle
 
\begin{abstract}
The thermally-pulsing asymptotic giant branch (TP-AGB) experienced by low- and intermediate-mass stars is one of the most uncertain phases of stellar evolution and the models need to be calibrated with the aid of observations.
To this purpose, we couple high-quality observations of resolved stars in the Small Magellanic Cloud (SMC) with detailed stellar population synthesis simulations computed with the \trilegal\ code.
The strength of our approach relies on the detailed spatially-resolved star formation history of the SMC, derived from the deep near-infrared photometry of the VISTA survey of the Magellanic Clouds, as well as on the capability to quickly and accurately explore a wide variety of parameters and effects with the \colibri\ code for the TP-AGB evolution.
Adopting a well-characterized set of observations -- star counts and luminosity functions -- we set up a calibration cycle along which we iteratively change  a few key parameters of the TP-AGB models  until we  eventually reach a good fit to the observations.
Our work leads to identify two best-fitting models that mainly differ in the efficiencies of the third dredge-up and mass loss in TP-AGB stars with initial masses larger than about 3~\Msun.
On the basis of these calibrated models we provide a full characterization of the TP-AGB stellar population in the SMC in terms of stellar parameters (initial masses, C/O ratios, carbon excess, mass-loss rates).
Extensive tables of isochrones including these improved models are publicly available.
\end{abstract}


\section{Introduction}
\label{sec:intro}

The thermally-pulsing asymptotic giant branch (TP-AGB) is an advanced stellar evolutionary phase during which stars of low- and intermediate-mass (with initial masses in the range $0.8~\Msun \la \Mini \la 6-8~\Msun$) experience helium and hydrogen double shell-burning, reach their highest luminosity, synthesise new chemical elements, undergo powerful mass loss via stellar winds, and then finally reach the stage of white dwarfs \citep{Herwig05}.
The impact of this evolutionary phase spans from chemical evolution and spectral energy distribution of galaxies to dust production in galaxies at low and high redshift
\citep{Maraston_etal_06, Conroy13, Zibetti_etal_13}.
Despite its importance in our understanding of galaxy evolution, TP-AGB modelling is still affected by large uncertainties due to the presence of several and interconnected processes -- i.e. third dredge-up (3DU), hot-bottom burning (HBB), stellar winds, long period pulsations, reprocessing of radiation by circumsttextllar dust -- for which a robust theory is still lacking \citep{Marigo_15}. 
 
Over the years, TP-AGB evolutionary models have been calibrated using primarily two different kinds of observations: the AGB stars in well-populated star clusters, and the luminosity functions of AGB stars in the fields of nearby galaxies. 

The first approach regards mainly the population of M- and C-type AGB stars found in the most populous star clusters in the Magellanic Clouds \citep{frogel90}, whose turn-off masses are generally well constrained, e.g. by means of isochrone fitting of deep colour-magnitude diagrams (CMDs). The 
observed numbers and luminosities of TP-AGB stars can be easily compared to the predictions from single-age, single-metallicity models, providing constraints to lifetimes, surface composition, and core masses of TP-AGB tracks with sub-solar metallicity \citep[e.g.][]{girardi07}, as a function of their initial masses.
This approach, however, is not completely satisfactory, and possibly fails, for a series of reasons. First, even in the most populous clusters the numbers of TP-AGB stars do not exceed two dozen,  and is typically much fewer and/or close to unity. These low number statistics lead to large uncertainties in all results that depend on the numbers of TP-AGB stars, including their lifetimes and their contribution to the integrated light of stellar populations. Second, due to the ``AGB boosting'' effect \citep{girardi13}, there is no guarantee that star counts in Magellanic Clouds' clusters of ages $\sim\!1.6$~Gyr -- the approximate age of the most populous ones -- are proportional to the lifetime in the TP-AGB phase. Finally, the discovery of broad main sequence turn-offs in some of the most massive Magellanic Clouds'
clusters, i.e. those with the largest early escape velocity \citep{goudfrooij15},
and their possible connection with fast rotation \citep{brandt15,goudfrooij17}, opens the possibility that the clusters' population may not be representative of the stars found in galaxy fields.

The second alternative is to use AGB stars in galaxy regions for which the distributions of ages and metallicities are well constrained, comparing their numbers and luminosities with those predicted through the stellar population synthesis approach. Given galaxy regions large enough to contain hundreds of AGB stars, this method overcomes the difficulties related to small-number statistics and the AGB-boosting in clusters. This technique was pioneered by \citet{groenewegen93}, who used the carbon star luminosity function in the Magellanic Clouds to calibrate the third dredge-up efficiency in synthetic AGB models. The same approach was later revised with increasingly detailed models of the AGB evolution \citep[e.g.][]{Marigo_etal99, marigo03, Izzard_etal_04, Stancliffe_etal_05, marigo13, marigo17}, and extended to many dwarf galaxies with good estimates of their star formation histories (SFHs) through analysis of Hubble Space Telescope (HST) photometry \citep[e.g.][]{girardi10,rosenfield14,rosenfield16}. The obvious shortcoming of this approach, compared to the use of star clusters mentioned above, is that we loose information about the initial stellar masses of TP-AGB stars. Indeed, although the main subtypes of TP-AGB stars broadly separate in CMDs like the $\ks$ versus \jks\ one, these sequences represent relatively wide mass ranges, in which tracks of different initial masses partially overlap. Despite the somewhat limited resolution in initial mass, the method benefits from a much better statistics.

Over the last decade, the quality of the data necessary to perform this calibration work using SMC field stars has improved significantly. First, catalogues of candidate TP-AGB stars were extended, from the initial near-infrared samples provided by the Deep Near Infrared Survey
of the Southern Sky \citep[DENIS,][]{denis} and the Two Micron All-Sky Survey \citep[2MASS][]{nikolaev00, cutri03}, to include the mid-infrared data from \textit{Spitzer} surveys \citep{gordon11, boyer11, SR16}; moreover, a non-negligible fraction of the SMC field TP-AGB stars now have a spectroscopic classification \citep{ruffle15,boyer15}.
Second, we have now a significantly better description of the space-resolved distribution of stellar ages, metallicities, distances, and mean extinctions across the SMC galaxy \citep{rubele15,rubele18}, as will be illustrated below. To these positive aspects, we can add the significant advancement in the physical prescriptions adopted in TP-AGB evolutionary models
\citep[e.g., gas opacities for varying chemical mixtures;][]{marigo_aringer09}, and the possibility of performing extremely fast calculations of extended model grids with the \colibri\ code \citep{marigo13}.

The combined progress of observations and models opens up the possibility of improving the calibration of the most uncertain processes in TP-AGB models. In this work, we use the population synthesis code \trilegal\ \citep{girardi05} to produce synthetic samples of TP-AGB stars in the SMC. The predicted star counts and luminosity functions (LFs) 
are compared to the observed ones from the AGB candidate list by \citet[][hereafter SR16]{SR16}. Our purpose is to put quantitative constraints on TP-AGB lifetimes as well as on the efficiency of the critical processes of mass loss and third dredge-up.

The structure of the paper is arranged as follows. 
We first describe the calibration strategy and the adopted sets of observations in Sect.~\ref{sec:calib_str}. Then, we recall the main properties of our TP-AGB models and their free parameters in Sect.~\ref{sec:models}. 
In Sect.~\ref{sec:calib_sets}  we introduce the large grid of new TP-AGB models and the procedure used to identify the best-fitting solutions. Section~\ref{sec:bestfit} is devoted to a discussion of the results and the implications derived from the calibration. A summary of the work and the main conclusions are outlined in Sects.~\ref{sec:sum} and~\ref{sec:concl}.

\section{Calibration strategy}
\label{sec:calib_str}

\begin{figure*}
  \includegraphics[width=\textwidth]{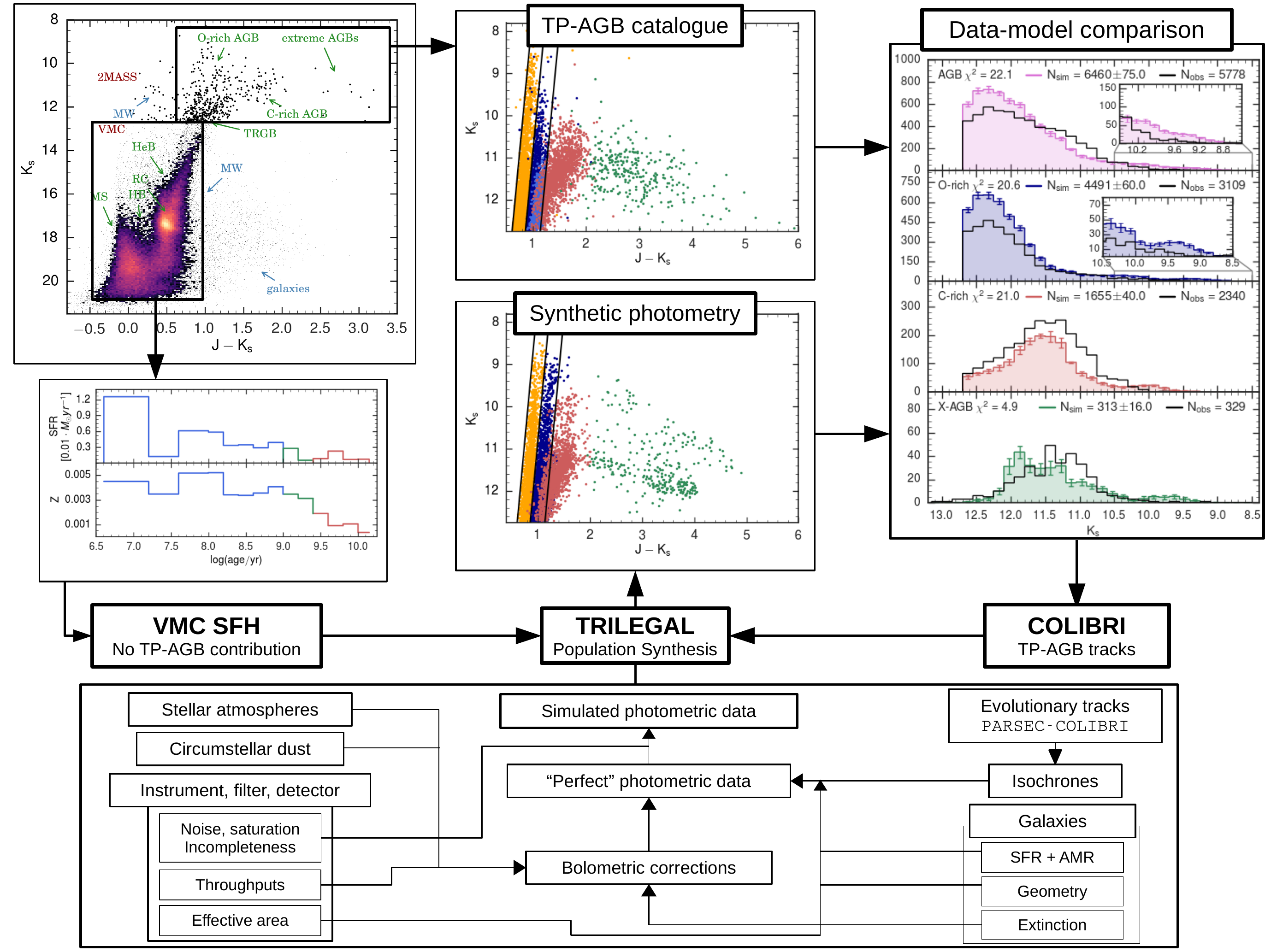}
  \caption{A scheme of the method and codes involved in this work. The \cmd{\ks}{J}{\ks} CMD at the top-left corner illustrates the
  observational data for one SMC subregion namely: (i) the deep VMC data \citep[see][and VizieR Catalog II/351/vmc\_dr4]{cioni11} which has allowed the derivation of a detailed spatially-resolved SFH for the SMC by \citet{rubele18}, at magnitudes not populated by TP-AGB stars, and (ii) the catalogues of TP-AGB stars built from the combination of 2MASS with mid-infrared surveys \citep{boyer11, SR16}. The central and right-hand panels illustrate the new steps performed in this work: The SFHs, together with TP-AGB evolutionary tracks from \colibri, are fed to the \trilegal\ code for the simulation of TP-AGB catalogues, for every subregion of the SMC. The simulations are then compared to the TP-AGB data (in terms of both star counts, luminosity functions and
  colour-magnitude diagrams), and new TP-AGB tracks are computed in response to the emerging discrepancies. Successive iterations then allow us to find the best-fitting models. The lower part of the figure describes the general working scheme of the \trilegal\ code.
  }\label{fig:scheme} 
\end{figure*}

The scheme in Fig.~\ref{fig:scheme} illustrates how our calibration cycle operates, and the data and codes  involved therein. A fundamental ingredient of our work is the spatially-resolved SFH of the SMC previously derived for many subregions of the SMC (Sect.~\ref{sec:SFH}). Using the SFH as an input, the \trilegal\ code is used to simulate the stellar populations of every subregion (Sect.~\ref{ssec:tri_sim}) that is also covered by the TP-AGB catalogues derived from 2MASS and \textit{Spitzer} observations (Sect.~\ref{sec:SMCdata}). These simulations are performed for different sets of TP-AGB models calculated with the \colibri\ code (varying a few key parameters, as detailed in Sect.~\ref{sec:models}), until a good overall description of the observed \ks-band LFs and \cmd{\ks}{J}{\ks} CMD
is reached (Sects.~\ref{sec:calib_sets} and~\ref{sec:bestfit}).

\subsection{Star Formation History}
\label{sec:SFH}

For many years, the most comprehensive study of the star formation history in the SMC was the spatially-resolved analysis performed by \citet{HZ04} for an area of 18~deg$^2$, based on the optical photometry from the Magellanic Clouds Photometric Survey (MCPS).  More recently, \citet[][hereafter R18]{rubele15,rubele18} recovered the SFH of the SMC using data from the VISTA near-infrared survey of the Magellanic Clouds \citep[VMC, ][]{cioni11}.
Both studies rely on CMD reconstruction methods. 
The main advantages of using VMC data are i) the lower extinction in the near-infrared passbands, and ii) the photometry reaching the oldest main-sequence turn-off points, ensuring a robust estimate of the ages. 

\begin{figure}
\centering
\includegraphics[width=\columnwidth]{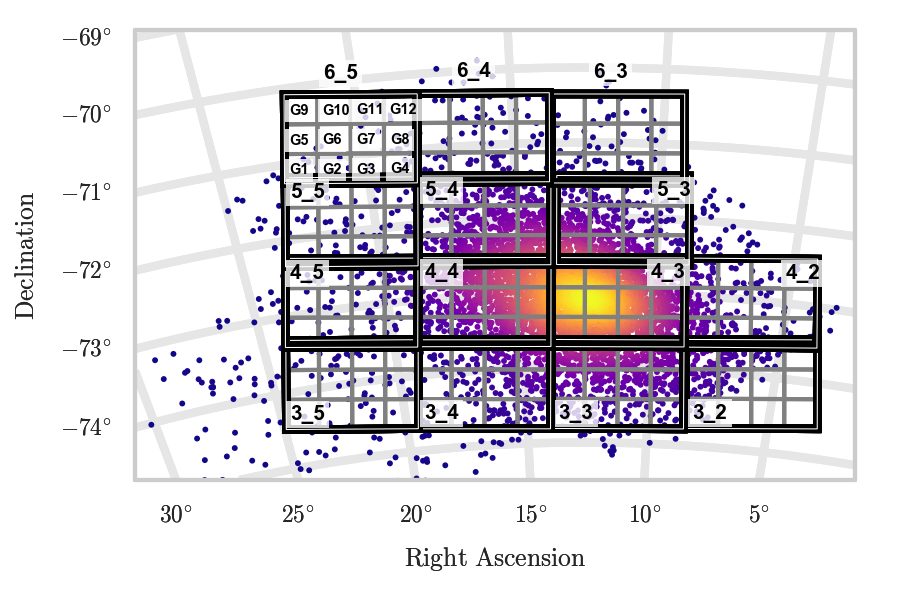}
\caption{VMC tiles used in the SFH recovery by R18. 
Each tile (in black) is subdivided into 12 subregions (in grey), as
illustrated for the tile 6\_5.
The background image shows the density map of the AGB stars classified by \citet{SR16}.}  
\label{fig:vmc_sage_map}
\end{figure}

In our study we use the SFH from R18. The most relevant aspects are:
\begin{enumerate}
\item The SFH recovery procedure was applied to 14 SMC tiles, each divided in 12 subregions of $0.143$~deg$^2$ each, for a total area of 23.57~deg$^2$, as illustrated in Fig.~\ref{fig:vmc_sage_map}, and listed in Table~\ref{tab:tab_tiles}.
\item The ``partial models'', representing the simple stellar populations used in the SFH recovery, were computed with an updated version of the \trilegal\ code \citep{marigo17}, which includes the most recent \parsec\ stellar evolutionary tracks\footnote{\url{http://stev.oapd.inaf.it/cmd}} \citep[v1.2S,][]{bressan12, bressan15} for evolutionary stages prior to the TP-AGB. The same sets of \parsec\ tracks are used in the present work, in addition to the TP-AGB tracks to be discussed later.
\item For each of the 168 subregions analysed, R18 independently derived the star formation rate (SFR), the age--metallicity relation (AMR), the mean distance modulus and the $V$-band extinction $A_{V}$, together with their $1\sigma$ uncertainties. 
\item The CMD regions used in the SFH recovery are limited to magnitudes $\ks>12.1$~mag and colours  $-0.52 <\jks\ \mathrm{[mag]}<0.88$ and  $-0.82 <\yks\ \mathrm{[mag]}<1.56$, therefore they include just a minor number of TP-AGB stars -- more specifically, the faint and bluest tails of their distributions in the CMD. Even so, those few TP-AGB stars are largely outnumbered by the RGB stars found in the same dataset at comparable magnitudes. This implies that the different TP-AGB tracks being tested in the present work, would not have affected the derivation of the underlying SFHs in any significant way.
\end{enumerate}

\begin{table} 
\caption{Central coordinates of the VMC tiles used in this work and number of AGB stars identified by SR16 for each tile.}
\label{tab:tab_tiles}
 
\begin{tabular}{ccccc}
\hline
\hline
Tile$^1$  & N. subreg.  & R.A.$_{\mathrm {J2000}}$ & Dec.$_{\mathrm {J2000}}$ & N. AGB \\
    & used & (deg)  & (deg.) &  \\
\hline
SMC 3\_2 &  4  & 5.8981   & $-$74.1159   & 136  \\
SMC 3\_3 & 11  & 11.2329  & $-$74.2117    & 578  \\
SMC 3\_4 & 11  & 16.588   & $-$74.1774    & 272  \\
SMC 3\_5 & 12  & 21.8784  & $-$74.0137    & 63   \\
SMC 4\_2 & 11  & 6.3087   & $-$73.0299    & 205  \\
SMC 4\_3 & 12  & 11.3112  & $-$73.1198    & 1888 \\
SMC 4\_4 & 12  & 16.3303  & $-$73.0876    & 1146 \\
SMC 4\_5 & 12  & 21.2959  & $-$72.9339    & 99   \\
SMC 5\_3 & 12  & 11.2043  & $-$72.0267    & 484  \\
SMC 5\_4 & 12  & 16.1088  & $-$71.9975    & 655  \\
SMC 5\_5 & 12  & 20.7706  & $-$71.8633    & 84   \\
SMC 6\_3 &  9 & 11.4532  & $-$70.9356    & 57   \\
SMC 6\_4 & 10 & 15.9581  & $-$70.8929    & 96   \\
SMC 6\_5 &  7 & 20.3437  & $-$70.7697    & 29   \\
\hline
 & N$_{\mathrm{Tot}}$  &  \multicolumn{2}{c}{Total}            &  \multicolumn{1}{c}{N$_{\mathrm{Tot}} $ }   \\
 & subreg. &  \multicolumn{2}{c}{area}   &  \multicolumn{1}{c}{AGB}   \\

&     147       &  \multicolumn{2}{c}{$\approx 21\sqdeg $}    & \multicolumn{1}{c}{5792}   \\
\hline
\multicolumn{5}{l}{ {\bf Notes:} } \\
\multicolumn{5}{l}{$^{(1)}$  Excluded subregions:}\\
\multicolumn{5}{l}{ \mbox{SMC 3\_2} G1, G2, G3, G4, G6, G7, G8, G12;}\\
\multicolumn{5}{l}{\mbox{SMC 3\_3} G1; \mbox{SMC 3\_4} G4; \mbox{SMC 4\_2} G4; \mbox{SMC 6\_3} G8, G11,}\\
\multicolumn{5}{l}{ G12; \mbox{SMC 6\_4} G6, G7; \mbox{SMC 6\_5} G5, G6, G9, G10, G11.}\\
\end{tabular}
\end{table} 

\subsection{Observations of AGB stars in the SMC}
\label{sec:SMCdata}

Our calibration is based on the SMC population of evolved stars identified and classified by \citet[][hereafter SR16]{SR16}, based on the program ``Surveying the Agents of Galaxy Evolution in the Tidally-stripped, Low Metallicity SMC'' \citep[SAGE-SMC,][]{gordon11}. The SAGE-SMC photometry includes optical to far-infrared passbands and covers the SMC main body (Bar and Wing) and tail regions. 
Fig.~\ref{fig:vmc_sage_map} shows the VMC survey coverage superimposed on the density map of the AGB stars classified by SR16. As can be seen, a large fraction of the area studied by R18 (about 87 per cent) is also covered by the SR16 catalogue.

The first catalogue of evolved stars based on the SAGE-SMC survey was published by \citet[][hereafter B11]{boyer11}, followed by the work of SR16 who reconstructed the candidate list using an updated version of the SAGE photometry and optical to mid-IR information from previous studies to refine the candidate selection. Details about their classification method are provided in the Appendix~\ref{app:classification}. In short, 
SMC stars in SR16 are classified into Red Super-Giant (RSG), Carbon-rich AGB (C-AGB or C-rich), Oxygen-rich AGB (O-AGB or O-rich), anomalous-AGB (a-AGB) and extreme AGB (X-AGB). The a-AGB star sample hosts both C- and O-rich stars that are thought to be low-mass evolved AGB stars \citep{boyer15}. The X-AGB sample contains the most dust-enshrouded stars and the majority of them are C-rich stars \citep[see e.g.][]{sloan06}. 
The classification of O-rich and C-rich stars is mainly based on their position in the \cmd{\ks}{J}{\ks} CMD,
where they draw distinct features. In addition, X-AGB and a-AGB stars are identified with the help of mid-infrared photometry.

The SR16 catalogue also includes the 81 sources spectroscopically classified by \citet{ruffle15}. In addition, we consider the spectroscopic classification of the 273 sources analysed by \citet{boyer15}. 
The star counts for the RSG/AGB populations are summarised in Table~\ref{tab:sage_stat} and they refer to the area covered by the VMC tiles shown in Fig.~\ref{fig:vmc_sage_map}. Only VMC subregions covered by the SR16 catalogue are considered. As in R18, we also exclude the subregion G6 of tile SMC 6\_4, which hosts the Milky Way globular cluster NGC~362. The number of AGB stars present in each tile according to the classification by SR16 is listed in Table~\ref{tab:tab_tiles}.  

\begin{table}
\centering
\caption{Number counts of RSG/AGB populations classified by SR16. The star counts refer to the selected SMC area used in this work. }
\label{tab:sage_stat}
\begin{tabular}{lc}
\hline
\hline
Population & Number of stars \\
\hline
C-AGB                & 1854     \\  
O-AGB                & 2623     \\ 
X-AGB                & 343      \\ 
a-AGB                & 972      \\  
RSG                  & 3150    \\ 
\hline
\end{tabular}
\end{table}
 
\subsection{TRILEGAL simulations}
\label{ssec:tri_sim}

Each subregion covered by both VMC and SAGE-SMC surveys is simulated according to its SFR, AMR, \av\ and distance as derived from the SFH recovery procedure. Most of the assumptions are identical to those adopted by R18:
\begin{itemize} 
\item [(i)] The adopted initial mass function (IMF) for single stars is that of \citet{kroupa01}.
\item [(ii)] Non-interacting binaries are simulated using a binary fraction of 30 per cent and a uniform distribution of mass ratios between the secondary and the primary, ranging from 0.7 and 1.
\item [(iii)] Our simulations are based on the same set of \parsec\ evolutionary tracks used to derive the SFH by R18, for all evolutionary phases prior to the TP-AGB. This ensures an optimal level of consistency between the two  works (see also Sect.~\ref{ssec:SFH_err}).  
\end{itemize}
For the purposes of the present study we build an extended grid 
of \colibri\ TP-AGB tracks, which are included in all simulations. The main steps for their integration in \trilegal\ are described in \citet{marigo17}, with some improvements to be detailed in Sect.~\ref{sec:models}. Our starting point is the set of TP-AGB stellar models described in \citet{rosenfield14, rosenfield16}, in which the mass loss in \colibri\ was adjusted to reproduce the observed TP-AGB luminosity functions in a sample of nearby galaxies imaged in the optical via the Advanced Camera for Surveys Nearby Galaxy Survey Treasury \citep[ANGST, ][]{Dalcanton_etal09}, and (for a subsample of them) in the near-infrared via a HST/SNAP follow-up campaign \citep{Dalcanton_etal12}.
These works pointed out the need of an efficient pre-dust mass loss, particularly for low-mass stars, in agreement with earlier findings \citep{girardi10}. Information about the third dredge-up could not be derived from those data, since the available HST photometry could not discriminate between O- and C-rich stars.

Model quantities are converted into the relevant photometry by means of extensive tables of bolometric corrections, derived from the stellar spectral libraries by \citet{aringer09} for C-rich stars, and \citet{castelli03} plus \citet{aringer16} for O-rich stars. 

The effect of the circumstellar dust in mass-losing stars is taken into account following the work of \citet{marigo08}, to which the reader should refer for all details. The approach couples radiative transfer calculations across dusty envelopes \citep{groenewegen06, bressan98}, with the scaling formalism introduced by \cite{ElitzurIvezic_01} and a few key relations from the dust-growth model originally developed by \citet{FerrarottiGail_06}. 
Tables of spectra are computed for a few dust mixtures and then interpolated to produce tables of bolometric corrections. The spectra used for the present simulations are computed for a fixed dust mixture consisting of amorphous carbon (85 per cent) and SiC (15 per cent) for C-rich stars,  and silicates for O-rich stars \citep{groenewegen06}
In this work we revise some input prescriptions of the dust treatment: we adjust the abundances of some elements (Si, S, Fe) to follow the scaled-solar pattern of \citet{caffau11} in line with the evolutionary tracks, and we compute the condensation degree of carbon dust based on the results from dynamical atmosphere models by \citet{Eriksson_etal14}, in place of a fitting relation based on the results of \citet{FerrarottiGail_06}. This modification improves the consistency of our simulations  since the same data presented in \citet{Eriksson_etal14} are also used to predict the mass-loss rates of C-stars during the dust-driven regime (see Sect.~\ref{ssec:mdot}).

To handle the effect of the AGB boosting period at $\sim\!1.6$~Gyr \citep{girardi13}, our simulations are split into three age intervals for each subregion (see Table~\ref{tab:age_res}). In the younger and older age intervals, the simulations are not affected by the boosting effect, therefore a resolution of 0.02~dex in log(age) suffices to represent the SFR for each subregion. During the boosting period, however, the log(age) resolution is improved to 0.001~dex, in order to sample the fast variations that occur in the production rate of TP-AGB stars. This boosting interval is intentionally kept very wide (0.4~dex),  to account for the variation of its mean age with metallicity. 

As a more general note, we recall that \trilegal\ is a quite complex population synthesis code able to generate catalogues covering stars over very wide ranges of stellar parameters, and including many physical processes specific to TP-AGB stars, like the temperature and luminosity variations driven by thermal pulses, changes in surface chemical surface and spectral type, and reprocessing of radiation by circumstellar dust shells \citep[see][]{marigo17}. However, there are two important effects still not considered in the present work. The first is the long-period variability that characterizes the photometry of most TP-AGB stars. Its effect is quite limited in the infrared filters considered here. Indeed, even Mira variables with large-amplitude variations in the optical have their $K$-band variations limited to just a few tenths of magnitude \citep[see figure 2.44 in][]{lattanzio_wood04}. The second potentially important effect is the evolution in interacting binaries, which depends on additional parameters like the distributions of orbital separations and mass ratios. While interacting binaries might be important to explain some classes of photometric variables and chemical anomalies in red giants \citep[see][for a review]{demarco17}, in the present work we assume that the fraction of stars that interact before the end of the TP-AGB evolution is negligible. This approach can be justified by the current stage of the TP-AGB model calibration, in which modellers still struggle to have uncertainties in model lifetimes reduced to within the $\lesssim\!20$ per cent level that characterizes other main evolutionary phases. 
Once this first step of the calibration is successfully completed, interacting binaries will have to be considered.

\begin{table} 
\centering
\caption{Age resolution in \trilegal\ simulations.}
\label{tab:age_res}
\begin{tabular}{lcc}
\hline
\hline
Period          & $\log(\mathrm{age/yr})$ interval & Resolution \\
                &       (dex)                      & (dex)       \\
 \hline
 Pre-boosting   &  6.600 - 9.000                   & 0.020 \\
 Boosting       &  9.000 - 9.400                   & 0.001 \\
 Post-boosting  &  9.400 - 10.170                  & 0.020 \\
 \hline
\end{tabular}
\end{table}

\subsection {SFH uncertainties}
\label{ssec:SFH_err}

First of all, the CMD reconstruction algorithm applied by R18 produces simulations that quite nicely fit the original VMC observations for every subregion of the SMC, as illustrated in their figure 4. Differences between star counts in the data and in their best-fitting models are in general $\la \!10$ per cent across the VMC colour-magnitude diagrams, and concentrated in the immediate vicinity of the red clump. Although such differences could be reflecting small uncertainties in stellar evolutionary models for pre-AGB phases, they are negligible compared to the sizeable heterogeneity in the TP-AGB lifetimes among various  models in the literature. Therefore, we can safely conclude that our simulations, using the same pre-AGB models and SFH as in R18, represent a robust starting point for the subsequent calibration of the TP-AGB models.

\begin{figure}
  \includegraphics[width=\columnwidth]{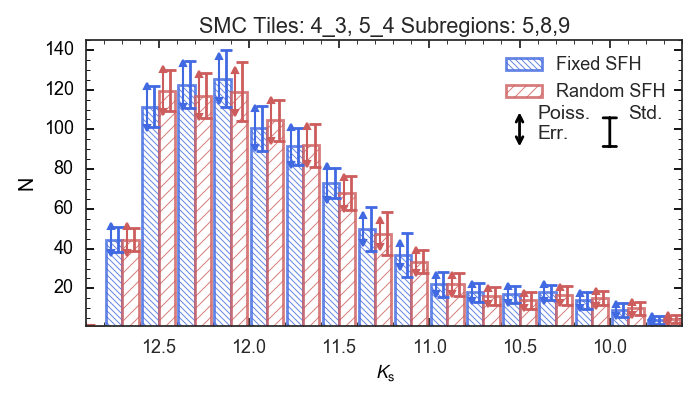}  
  \caption{Comparison between the mean \ks-band LFs derived from two sets of simulations of 6 SMC subregions, derived from the best-fitting SFH values (blue histograms), and from randomly sampled SFH (red histograms). The two sets of error bars illustrate the standard deviation, calculated from the 10 \trilegal\ simulations, and the Poisson error, i.e. the square root of the number counts, for each \ks\ bin. 
}\label{fig:SFHerr1}
\end{figure}

We verify how the uncertainties in the best-fitting SFR, AMR, distance
and extinction impact the simulated LFs and number counts. To this aim,
we select the subregions 5, 8, and 9 of the SMC tiles 4\_3 and 5\_4,
which can be considered as representative of most of the SMC regions.
For each subregion, we compute two distinct sets of 10 simulations.
The first set is computed using the best-fitting values of the SFH.
For the second one, the value of the SFH at each age bin is
extracted from a Gaussian distribution, centred at the best-fitting value,
and with standard deviation equal to the uncertainty associated
with the best-fitting solutions\footnote{
We use a single Gaussian distribution to describe the symmetric uncertainties in
the values of distance and reddening. To describe the asymmetric uncertainties
in the SFR and AMR, we use two Gaussian distributions, having the same
mean but different standard deviations, $\sigma_{\rm low}$ and $\sigma_{\rm up}$.
}.

Figure~\ref{fig:SFHerr1} compares the mean \ks-band LFs for the two sets of simulations.
We see that: (i) the standard deviation and the Poisson error in each luminosity bin are equivalent, both in the case of a fixed SFH (as expected) and in the case with randomly extracted SFH, and (ii) the difference in number counts between the two cases is within the Poisson error in each luminosity bin. Additionally, we find that the uncertainties in the SFH produce smaller than 10 per cent variations in the total number of AGB stars. These results justify the use of a simplified approach, in which, for each SMC subregion, the LF and the associated uncertainty
in each luminosity bin are obtained as the mean value and the standard deviation from 10 \trilegal\ simulations obtained with the best-fitting values of the SFH.
Given the large number of TP-AGB sets we tested, this approach
allows us to significantly speed-up the calibration procedure.

\subsection{Model selection criteria}
To compare our results to the observed catalogue, we need to properly identify the different classes of stars in the synthetic catalogue.
One possibility is to use the same photometric criteria as in B11, also applied in SR16. In this case, however, small offsets in the model colours and magnitudes with respect to the observations can significantly mix different kinds of stars, affecting the comparison with the observed star counts in the different AGB populations. On the other hand, the catalogues of synthetic stars contain the surface chemical composition that allows the straightforward separation into different chemical types.
For these reasons, we used a `hybrid' selection approach: whenever possible we use theoretical parameters -- i.e.\ C/O ratio, evolutionary stage -- to distinguish different populations, complementing the selection with photometric criteria when necessary (i.e. for the X-AGB stars). Our selection criteria are detailed in Appendix~\ref{ssec:sel_criteria}. 

In addition, we have identified a systematic colour shift between models and observations, which appears to widen at increasing luminosity of the O-rich stars (including RSG and O-AGB stars). In Appendix~\ref{app:colorshift} we discuss this feature and describe the way models are corrected to reduce its impact on the comparison between data and simulations.

\subsection{Metrics for identifying the best-fitting model}
\label{subsec:metrics}
 
Before proceeding with the model calibration, we should better establish what exactly defines a good TP-AGB model in our case: essentially, it means producing the correct lifetimes and luminosity distributions for the main sub-types of TP-AGB stars. In this perspective, we use as primary calibrators the star counts in the \cmd{\ks}{J}{\ks} CMD and the \ks-band LFs. The basic requirement is that the models simultaneously reproduce these quantities for the O-, C-, and X-AGB star samples identified by SR16. Among the many possible criteria to quantify the performance of each simulation, we choose the  $\chi^2$ of the LF distributions,   \chisqlf.  It is computed following the $\chi^{2}$-like statistics defined by \citet{dolphin02}, i.e.\ assuming that the stars are distributed into Hess diagram cells following a Poisson statistics: 
\begin{equation}
\chisqlf =\frac{1}{N} \sum_{i=1}^{N} 2\cdot \left[ m_{i} - n_{i} + n_{i}\cdot \ln \left(\frac{n_{i}}{m_{i}}\right) \right ]
\end{equation}
where $n$ and $m$ are the number of simulated and observed stars, respectively, in the $i$-th cell of the Hess diagram or the $i$-th bin of the LF, and $N$ is the number of cells or bins in which both $n$ and $m$ are different from zero.

For each of the three classes of AGB stars and for the entire sample, we compute the value of the \chisqlf. The performance of each set of models is evaluated using a total $\overline{\chi^{\,2}_{\rm \,LF}}$, that is the average of the four \chisqlf\ values weighted by the respective number of observed stars.
To identify the best-fitting model, we first select the sets that simultaneously reproduce the total number of observed AGB, O-, C- and X-AGB within 3-$\sigma$. The one with the lowest $\overline{\chi^{\,2}_{\rm \,LF}}$ value is considered as the best-fitting solution. 
This criterion is applied to the two series of models presented in Sects.~\ref{ssec:changemdot} and~\ref{ssec:changedup}.
 

\section{TP-AGB models}
\label{sec:models}

The evolution prior to the TP-AGB phase is computed with the \parsec\ code \citep{bressan12}, while the TP-AGB phase is followed by the \colibri\ code \citep{marigo13}. We refer the reader to this latter paper for a detailed explanation of all TP-AGB calculations. Below we mention just a few details, aimed to provide a quick picture of what is being considered.

\subsection{The \parsec\ and \colibri\ codes: basic assumptions}

\parsec\ v1.2S models take into account overshooting from stellar cores, following the \citet{bressan81} formalism. The overshooting parameter, $\Lambda_\mathrm{ov}$, is assumed to increase from null to its maximum value, 0.5 pressure scale heights ($H_P$; measured across the convective border), over a 0.3-\Msun\ wide interval that starts at the maximum mass of stars with radiative cores on their main sequence (which is 0.95 to 1.1~\Msun, depending on metallicity). This maximum value of overshooting implies an overshooting region of $\sim0.25\,H_P$ above the convective border (and, equivalently, a $f_\mathrm{ov}\sim0.022$ parameter in diffusive methods), which is among the typical values in the current literature \citep[see][and references therein]{claret16,claret17}. This overshooting efficiency determines the maximum masses of the TP-AGB models, referred to as $M_\mathrm{IM}$ in \citet{bressan12}. In practice, \parsec\ models of smaller masses develop at least one thermal pulse in the helium shell, after forming an elecron-degenerate carbon-oxygen core. The \colibri\ evolution starts at this point. The $M_\mathrm{IM}$ limit depends on metallicity and is generally located between 5 and 6~\Msun\ \citep[see figure 3 in][]{bressan12}. Stars with slightly higher masses (up to $\sim9$~\Msun, in models with overshooting), build a degenerate oxygen-neon core before developing any significant thermal pulse. These stars are expected to evolve through a short-lived super-AGB phase \citep[e.g.][]{Siess_2010}, and will not be considered in this paper.

\parsec\ models are  computed assuming the mixing length theory with an $\alpha_\mathrm{MLT}=1.74$ parameter, as determined from the calibration of the solar model \citep{bressan12}. \colibri\ adopts the same description for the external convection. The physical process most affected by this choice is the HBB, whose energetics and nucleosynthesis are computed by means of a detailed nuclear network \citep{marigo13}. In Sect. \ref{sec:HBBimpact} below will discuss the impact of this model assumption in our calibration.

\colibri\ includes the \texttt{\AE SOPUS} code \citep{marigo_aringer09} as an internal routine, to compute the equation of state on-the-fly (for $\simeq 800$ species) and Rosseland mean opacities in the outer regions of the star. Moreover, it accounts for the hot-bottom burning nucleosynthesis and energetics through a complete nuclear network coupled to a diffusive description of convection. 
Given the uncertainties in the mass loss and 3DU, these processes are treated with parametrized descriptions. The related free parameters need to be calibrated to reproduce the observations (see Sects.~\ref{ssec:changemdot} and \ref{ssec:changedup}).
For any combination of input prescriptions, the code allows for a very quick computation of large grids of TP-AGB tracks. Such a feature makes \colibri\ a suitable tool for the calibration cycle depicted in Fig.~\ref{fig:scheme}.

 \colibri\ usually starts from a stellar configuration (defined by luminosity, total mass,  core mass, and envelope composition) extracted from a \parsec\ model just before its first significant thermal pulse (1TP). The subsequent TP-AGB evolutionary track is then joined with the \parsec\ one, inside \trilegal, to form a continuous track spanning from the pre main-sequence to the end of the TP-AGB. As illustrated in \citet{marigo13}, there is in general a quite good continuity in evolutionary properties as a given track moves from \parsec\ to \colibri. 

 However, \parsec\ tracks are computed at constant mass, whereas in this work we allow for the possibility that mass loss occurs before the first thermal pulse. Therefore, we adopt an additional step before joining \parsec\ and \colibri\ tracks: along the final sections of the \parsec\ tracks (corresponding to Early-AGB stages), we compute the mass-loss rate according to the adopted mass-loss formalism. At each time step, the effective temperature along the \parsec\ track is re-evaluated, using the envelope integration routines in the \colibri\ code, together with the corrected stellar mass. 
 The entire procedure is justified by the fact that, as the TP-AGB is approached, the 
 luminosity and the rate of brightening are mainly controlled by the core mass, and are much less sensitive to the envelope mass. The effect of accounting for mass loss during the Early-AGB is discussed in Sect.~\ref{ssec:changemdot}.
 
In the following we outline the main ingredients that are relevant to the physical calibration.

\subsection{Mass loss}
\label{ssec:mdot}

The treatment of mass loss in AGB models is critical as it controls the duration of the phase, which normally terminates when the whole envelope is ejected into the interstellar medium. 
The abundant literature on the topic proves that over the years substantial efforts have been devoted to understand the physical mechanisms responsible for mass loss in AGB stars, and more generally in red giants \citep[see][for extensive reviews]{HoefnerOlofsson_18, Willson_00}.
While the key roles of stellar pulsation and dust growth are today universally acknowledged,  a comprehensive theoretical framework is still missing. 
As has been assumed in earlier studies \citep{girardi10, rosenfield14, rosenfield16}, we adopt a scheme that considers two regimes of mass loss, which we shortly refer to as pre-dust mass loss (with a rate \mdotpre) and dust-driven mass loss (with a rate \mdotdust), respectively. 

In our scheme, the pre-dust mass loss occurs during red-giant stages,  characterized by relatively low luminosities and high effective temperatures, which set unfavourable conditions to drive a stellar wind by radiation pressure on dust grains. 
During these phases mass loss should instead be produced by another, still not clearly identified, mechanism. In this work we implement a routine based on the model developed by \citet[][hereafter CS11]{CranmerSaar_11}, in which the wind is assumed to be driven by Alfv\'en waves and turbulence that originate in cool extended chromospheres \citep[see also][]{SC05}. 
To this aim we adopt all the equations described in the paper by \citet{CranmerSaar_11} related to
the cold wave-driven mass loss case (their section 3.2), except for the photospheric parameters
(e.g. opacity, mass density, gas pressure) that are obtained directly from the \colibri\ code.

At later stages along the AGB, we assume that the chromospheric mass loss is quenched and the outflow is initially triggered through stellar pulsations \citep{McDonald_etal_18} and then fully accelerated by radiation pressure on dust grains thanks to the dust--gas dynamical coupling \citep{HoefnerOlofsson_18}.
Several relations for \mdotdust\ as a function of stellar parameters (current mass $M$, luminosity $L$, effective temperature \Teff, initial metallicity $\Zini$) are available in the literature.
Here we investigate a few among the most popular ones, i.e. \citet[][hereafter VW93]{VassiliadisWood_93}, which relates the efficiency of mass loss to the pulsation period\footnote{The pulsation periods are computed as in the original paper of VW93.}, and \citet[][hereafter BL95]{bloecker95}, which is characterized by a significant dependence on the luminosity. In addition, we use  the recent results of dynamical atmosphere models for carbon stars \citep{mattsson10, Eriksson_etal14} to
predict $\dot M_{\rm dust}$ as a function of stellar parameters: mass, luminosity, effective temperature, and carbon excess, \cminuso. This latter is defined as:
\begin{equation}
\cminuso = \log (n_{\rm C} - n_{\rm O} ) - \log(n_{\rm H}) + 12 \, ,
\end{equation}
where $n_{\rm C}$, $n_{\rm O}$, and $n_{\rm H}$ denote the number densities of carbon, oxygen and hydrogen, respectively.
During the TP-AGB evolutionary calculations, if not otherwise specified, the current mass-loss rate is taken as:
\begin{equation}
\label{eq:cur_mdot}
\dot M = \mathrm{max}(\etapre \cdot \mdotpre, \, \etadust \cdot \mdotdust) \,,
\end{equation}
where \etapre\ and \etadust\ 
are adjustable efficiency parameters.

Mass loss on the RGB of low-mass stars, instead, is taken into account by simply removing the amount of mass inferred from the \citet{reimers} formula with a multiplicative factor $\eta_\mathrm{Rei}=0.2$ \citep[see][]{miglio12}, as we pass from the RGB-tip to the initial stage of core-helium burning. At SMC metallicities, only stars with $\Mini\lesssim1.2$~\Msun\ turn out to have their RGB masses reduced by more than 5 per cent. Therefore, RGB mass loss does not impact on the TP-AGB populations here simulated in any significant way.
   
\subsection{The third dredge-up}
\label{ssec:3dup}
Given the uncertainties and the heterogeneous results that characterize convective mixing in full TP-AGB models, we need to  treat the third dredge-up following a parametric scheme as detailed in \citet{marigo13}. We need to specify three main characteristics of the third dredge-up: i) the onset, ii) the efficiency, and iii) the chemical composition of the intershell material. We now discuss each of these prescriptions in turn.

\subsubsection{Onset of the 3DU}
Following earlier studies \citep{Marigo_etal99, marigo07},  in \colibri\ the occurrence of a mixing event is determined by a temperature criterion based on the parameter $\Tbdred$, which is the minimum temperature that should must be reached at the base of the convective envelope  for a mixing event to take place. Typical values lie in the range $\mathrm{6.3 \la \log(\Tbdred/\mathrm{K}) \la 6.7}$, as first shown by \citet[][]{Wood_81}. At each thermal pulse, the \colibri\ code performs complete envelope integrations to check whether the temperature criterion is satisfied or not at the stage of the post-flash luminosity maximum.
An alternative choice, equivalent to $\Tbdred$, is the classical parameter $\Mcmin$ that defines the minimum core mass for the onset of the sequence of third dredge-up events. Typical values are usually within the range $0.54 \la \Mcmin\, (\Msun) \la 0.60 $. 

In this work we rely on the $\Tbdred$ parameter. The simplest approach is to adopt a constant value
(e.g. $\log(\Tbdred/\mathrm{K}) = 6.4$) in all models. This assumption may be relaxed to include a dependence on metallicity,  so that the minimum temperature increases with $\Zini$, as indicated by \citet{marigo07} following the results of their calibration.  The effect is to increase the minimum initial mass for the formation of carbon stars at higher metallicity.  In our models we test both the case of constant $\Tbdred$ as well as metallicity-dependent relations. 

\subsubsection{Efficiency of the 3DU}

The efficiency of each mixing event is usually described by
the parameter $\lambda = \Delta M_{\rm dred}/\Delta M_{\rm c}$,
the fraction of the increment of the core mass during an
inter-pulse period that is dredged up during the next thermal pulse. 
Typical values are in the range $\lambda=0$ (no dredge-up)
and $\lambda=1$ (no net increase of the core mass), even though
cases with $\lambda > 1$ are predicted in models that assume
efficient convective overshoot \citep[e.g.][]{Herwig_04}. 

\begin{figure}
\centering
    \includegraphics[width=\columnwidth]{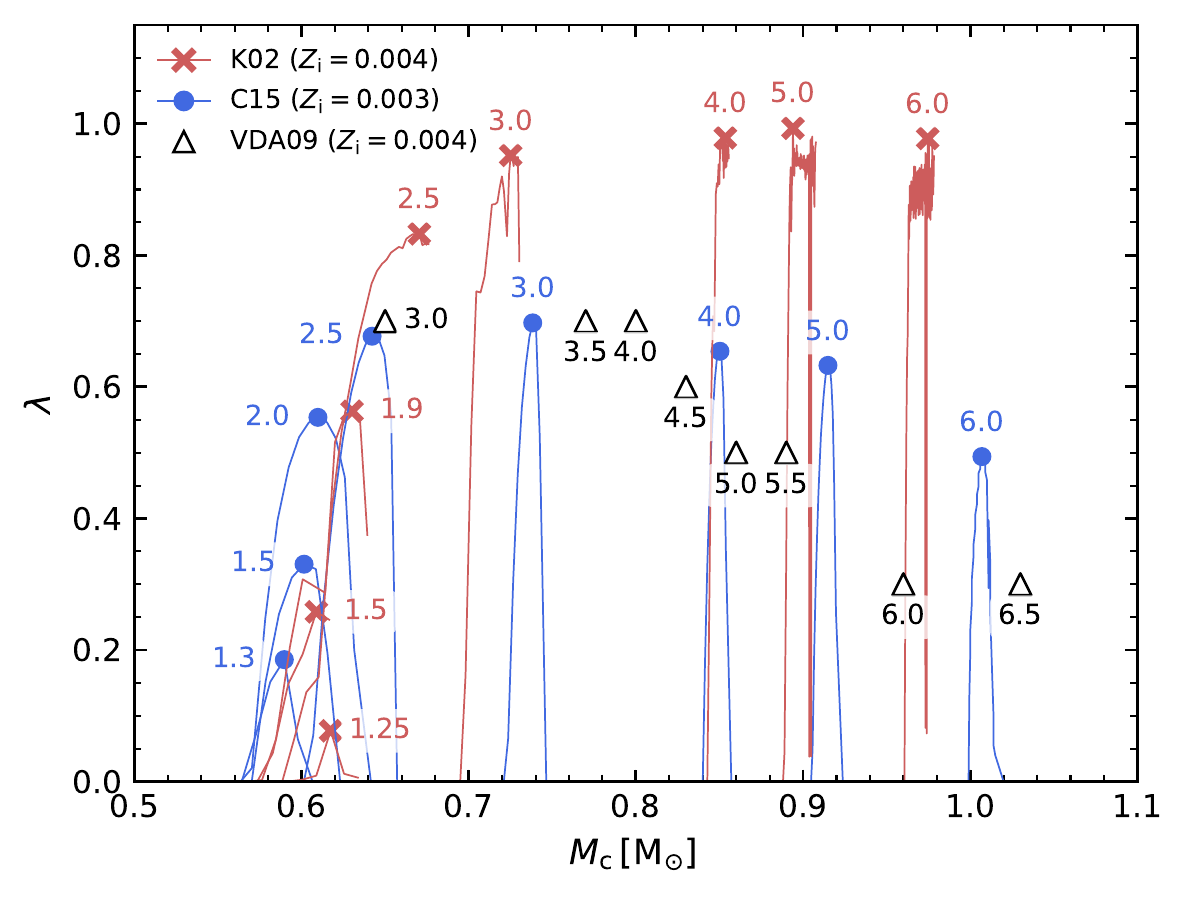}
    \caption{
    Predictions of the maximum 3DU efficiency $\lambda_{\rm max}$ as a function of the core mass from a few complete TP-AGB models with initial metallicity  $\Zini = 0.003, 0.004$ and initial mass as indicated.  
    The predicted values of
    $\lambda_{\rm max}$ from \citet[][K02]{karakas02} and \citet[][C15]{Cristallo_etal15} are shown with red crosses and blue filled circles, respectively. The predicted values of
    $\lambda_{\rm max}$ as a function of the core mass at the first thermal pulse from the models of \citet[][VDA09]{VenturaDantona_09} are shown with empty triangles. 
    }
    \label{fig:fruity_lamb}
\end{figure}

We aim to devise a parametric description of $\lambda$
that a) is qualitatively consistent with full TP-AGB models calculations,
and b) keeps the necessary flexibility,
through the adoption of free parameters, to perform a
physically-sound calibration based on observations. Fig.~\ref{fig:fruity_lamb} provides an example of the
3DU efficiency predicted by complete TP-AGB models from a few different authors.
During the TP-AGB evolution of a given star,
$\lambda$ first increases following the strengthening of thermal pulses,
reaches a maximum value $\lambda_{\rm max}$, and then decreases
as the envelope mass is reduced by stellar winds, until the
3DU is eventually quenched.
The peak value $\lambda_{\rm max}$ is not fixed,
but varies from star to star depending on the initial stellar mass.
We note, however, that this behaviour is not homogeneously
described by existing models.
For instance, the models of \citet[][hereinafter also K02]{karakas02},
and, in general, all TP-AGB models computed with the Mount Stromlo Evolutionary code
\citep[e.g. ][and references therein]{karakas02, Karakas_10, Fishlock_etal_14},
reach a high 3DU efficiency ($\lambda\simeq1$) for $M_{\rm i}\gtrsim3\,{\rm M}_{\odot}$,
regardless of metallicity.
In contrast, TP-AGB models by \citet[][the FRUITY database]{Cristallo_etal_11, Cristallo_etal15}
and by \citet{VenturaDantona_09} are characterized by lower values of $\lambda$,
and by a decreasing trend of $\lambda_{\rm max}$ with the initial stellar mass.

\begin{figure}
\centering
    \includegraphics[width=0.85\columnwidth]{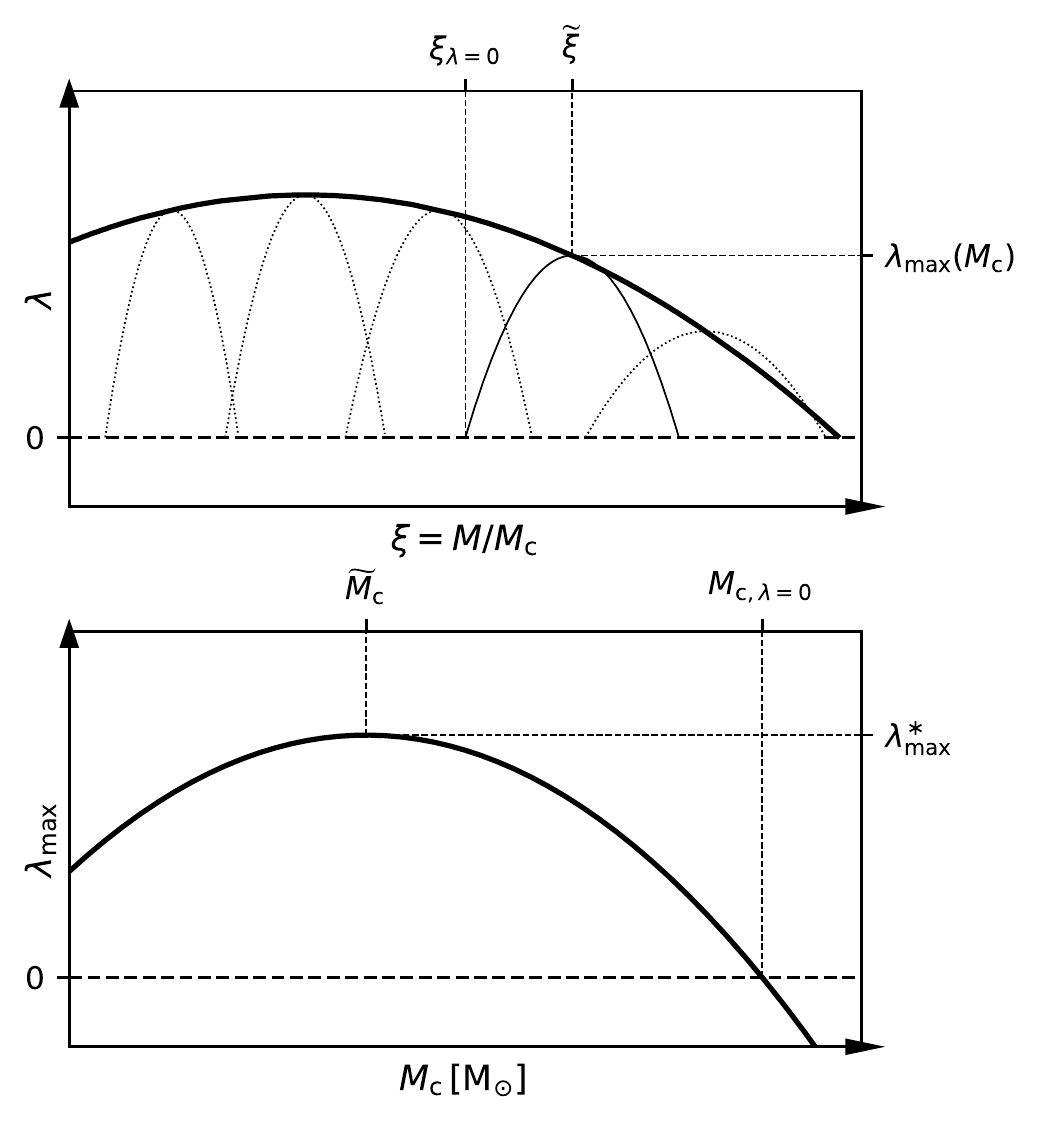}
    \caption{ Schematic depiction of the meaning of the free parameters in our
    new formalism for the 3DU efficiency. 
    Top panel: evolution of $\lambda$ as a function of $\xi$ for
    individual models having different values of the core mass (thin lines).
    For each model, $\lambda$ reaches a maximum $\lambda_{\rm max}$ at $\widetilde{\xi}$ and drops to zero at $\xi=\xi_{\lambda=0}$, representing the quenching of the 3DU when
    the envelope mass is substantially reduced by mass loss. The maximum value of $\lambda$ for each model is assumed to be a function of the core mass (thick line). Bottom panel: $\lambda_{\rm max}$
    increases with core mass to a peak $\lambda_{\rm max}^{\ast}$,
    the maximum 3DU efficiency among all TP-AGB models, met at $M_{\rm c}=\widetilde{M}_{\rm c}$,
    then decreases until $\lambda_{\rm max}=0$ at $M_{{\rm c},\lambda=0}$,
    beyond which the 3DU does not occur. }
    \label{fig:lambda_scheme}
\end{figure}

To account for this behaviour of the 3DU efficiency,
we express $\lambda$ as a function of the ratio between the
current total mass and the current core mass, $\xi=M/M_{\rm c}$,
a dimensionless parameter that is expected to decrease during the TP-AGB evolution
as a consequence of mass loss and core mass growth.
A scheme of the new parametric formalism is shown in Fig.~\ref{fig:lambda_scheme}. 
We assume a parabolic dependence for $\xi$, that allows to capture the bell-shaped evolution of $\lambda$.
The controlling parameters for this relation are
the centre $\widetilde{\xi}$ of the parabola, the value
$\xi_{\lambda=0}$ at which $\lambda$ drops to zero,
and the peak value $\lambda_{\rm max}$.
The latter is not treated as a free parameter, but rather
expressed as a parabolic function $\lambda_{\rm max}(M_{\rm c})$ of the core mass,
to reflect the increase in 3DU efficiency with initial stellar mass,
while also allowing for a decrease towards the high-mass range.
The controlling parameters, equivalent to the previous case,
are $\widetilde{M}_{\rm c}$, $M_{{\rm c},\lambda=0}$, and
$\lambda_{\rm max}^{\ast}$.

The explicit form of our parametrization is:
\begin{equation}
    \lambda = \lambda_{\rm max}^{\ast}
        \left[\left(\frac{\xi-\widetilde{\xi}}{\widetilde{\xi}-\xi_{\lambda=0}}\right)^2-1\right]
        \left[\left(\frac{M_{\rm c}-\widetilde{M}_{\rm c}}{\widetilde{M}_{\rm c}-M_{{\rm c},\lambda=0}}\right)^2-1\right]
        \,,
\end{equation}
that is used to assign the 3DU efficiency at each thermal pulse
as a function of $M$ and $M_{\rm c}$, provided the $T_{\rm b}^{\rm dred}$ criterion for the occurrence of the 3DU is fulfilled.
The physical interpretation of the parameters is the following.
\begin{enumerate}
    \item  For a given star, $\widetilde{\xi}$ is the value of $M/M_{\rm c}$ at which the maximum $\lambda$ is reached, while $\xi_{\lambda=0}$
    is the value at which the 3DU is quenched due the decrease of the
    envelope mass caused by mass loss. The adopted ranges for these parameters ($\widetilde{\xi}\simeq3$~--~$4$, $\xi_{\lambda=0}\simeq1.5$~--~$2$)
    are based on full evolutionary calculations available in the FRUITY database \citep{Cristallo_etal15}.
    \item $\lambda_{\rm max}^{\ast}$ is the maximum efficiency of the 3DU
    among all TP-AGB stars, that is attained at $M_{\rm c}=\widetilde{M}_{\rm c}$,
    while $M_{{\rm c},\lambda=0}$ is the value of core mass beyond which
    the 3DU does not occur. These are the key parameters that we aim at
    constraining with the aid of the observed AGB stars in the SMC.
\end{enumerate}

\subsubsection{Chemical composition of the interhsell}
In the \colibri\ code the pulse-driven nucleosynthesis is computed coupling a synthetic description of the intershell with a complete nuclear network which includes the most important $\alpha$-capture reactions. In addition, we have the possibility to explore the impact of various efficiencies of the convective overshoot \citep[refer to Sect. 7.5.5 in][]{marigo13}. In the present work we limit to consider the standard case for the intershell composition (i.e. no convective overshoot, oxygen-poor case), with typical abundances (in mass fraction) of helium, carbon and oxygen of  $\mathrm{^4He/^{12}C/^{16}O \approx 0.70-0.75/0.25-0.20/0.005-0.01}$.
This pattern may be significantly altered by the inclusion of convective overshoot at the bottom of the pulse-driven convective zone \citep{Herwig_04}, the main effect being the enhancement of the oxygen abundance.
We plan to extend our investigation about the impact of an oxygen-rich intershell to a future study. 


\section{Calibration along the sequence of TP-AGB sets}
\label{sec:calib_sets}

In this section we first describe the results of our starting model. We then move to illustrate our calibration presenting the large grid of computed models and their performance  in comparison with observations.
In a first series of models we explore the effect of different mass-loss prescriptions for both O- and C-rich stars. Starting from the resulting best fitting-model of this series, we calculate a second series of models in which we focus on the calibration of the 3DU.

\subsection{The starting set of TP-AGB models}
\label{subsec:starting}

As anticipated in Sect. \ref{sec:calib_str}, we start by simulating the SMC population using the TP-AGB tracks described in \citet{rosenfield14, rosenfield16} and presented in \citet{marigo17}.
We will refer to this initial set as S\_00.  The mass-loss prescription for the pre-dust driven wind is a modified version of the \citet[][hereafter SC05]{SC05} prescription \citep{rosenfield16}.
The dust-driven phase is described with a mass-loss formalism similar to \citet{bedijn88}, for both O-rich and C-rich stars. The 3DU onset is controlled by a constant temperature parameter $\log(\Tbdred/\mathrm{K}) = 6.40$ and the efficiency parameter $\lambda$ follows the relations provided by K02.

\begin{figure}
\centering
\includegraphics[width=0.9\columnwidth]{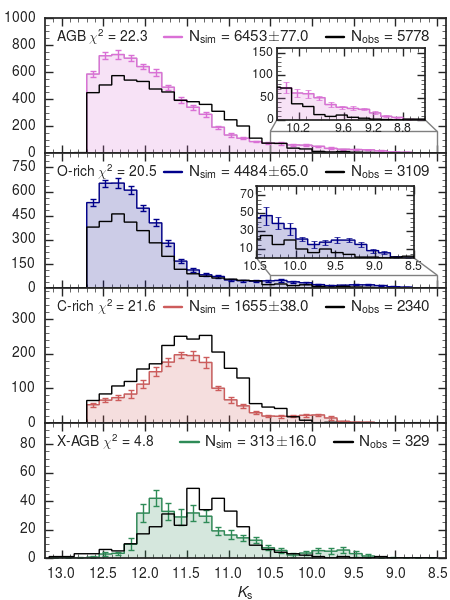}
\caption{Mean \ks-band LFs from S\_00 models (filled histograms in colour), as compared to the observations (dark-line histogram), both for the entire TP-AGB sample (top panel) and for the three main classes of TP-AGB stars (other panels). The error bars cover the 1-$\sigma$ standard deviation of the 10 \trilegal\ realisations.  The  total number of observed stars,  synthetic stars, and the \chisqlf\ specific to each simulated LF are also indicated.}
\label{fig:LFKsS000}
\end{figure}

Fig.~\ref{fig:LFKsS000} compares the observed \ks-band LFs for each class of AGB stars with the simulated ones based on the set S\_00. 
As shown in the top panel, the total number of AGB stars is reproduced to within 10 per cent, a result that supports the earlier calibration performed by \citet{rosenfield16}. However, from a closer inspection  we note that the model underpredicts the number of stars at magnitudes brighter than $\ks \approx 11.6$~mag, and overpredicts the number of stars at fainter magnitudes.
The deficit of predicted stars at brighter magnitudes corresponds mainly to an underestimation ($\approx 30$ per cent) of C-rich stars (third panel from top).
The excess of predicted stars at fainter magnitudes corresponds to an overestimation ($\approx 40$ per cent) of O-rich AGB stars with initial stellar mass in the range 1--1.5~\Msun\ and low metallicity, $\Zini\approx 0.001$ (second panel from top).
A slight excess is also present at the bright end of the LF of O-rich stars. Our analysis  indicates that such stars are experiencing HBB, with  initial masses larger than $\approx$ 3~\Msun\ and initial metallicity $\Zini \simeq$ 0.003--0.004.

The total number of X-AGB stars is in agreement with the observations within $\approx 15$ per cent (bottom panel), but the shape of the LFs shows discrepancies around  $\ks \approx 12$ mag and  $\ks \approx 11$ mag. 
In particular the peak of the simulated LF is located at fainter $\ks$ magnitudes compared to the observed one.
The difference is likely a consequence of the adoption of the \citet{VassiliadisWood_93} prescription for the mass loss during the super-wind phase. With that prescription, mass-loss rates reach a maximum ($10^{-5}$--$10^{-4}~\Msun \rm{yr}^{-1}$) and then remain almost constant during the final thermal pulses, causing  significant dust extinction in the $\ks$-band and an accumulation of stars at $\jks \ga 4$ mag and $\ks \approx 12$~mag.
 
\begin{figure}
\centering
\includegraphics[width=\columnwidth]{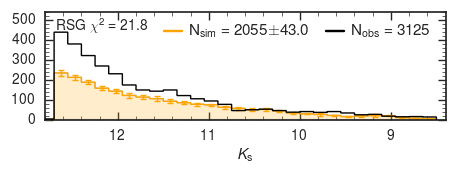}
\caption{Similarly to Fig.~\ref{fig:LFKsS000}, the \ks-band LFs for the RSGs.}
\label{fig:LFKsS000_RSG}
\end{figure} 
 
Before leaving this section we consider it worth commenting on the sample of red super-giants present in the SMC data. 
The comparison between observed and simulated LFs is shown in  Fig.~\ref{fig:LFKsS000_RSG}.
The overall shape of the observed LF of the RSG stars is well reproduced by the starting  model. However, there is an evident lack of stars at faint magnitudes. Such discrepancy, already pointed out by \citet{melbourne12}, can be due to a number of reasons: (i) a residual foreground contamination in the data up to $\approx$ 35 per cent \citep[][their Sect. 2.2]{boyer11}; (ii) an underestimation of the SFR for these young populations; (iii) an underestimation of predicted lifetimes for core-helium burning stars with intermediate and high initial masses. Since the present calibration does not rely on this class of stars, we will only focus our analysis on AGB stars, postponing the investigation of the RSGs to a future work.

\subsection{The new grid of TP-AGB tracks}
Starting from the results obtained with the initial S\_00 set,  we put in operation a calibration cycle along which the mass-loss prescriptions and the third-dredge up parameters are varied until a satisfactory reproduction of the observations is achieved.
The calibration procedure relies on a large grid of TP-AGB tracks computed with \colibri, adopting different descriptions for the third dredge-up and mass loss.
Each combination of the adopted parameters corresponds to a set of tracks. Each set covers a range of initial masses from about 0.5~\Msun\ to 5--6~\Msun\ (typically 70 values) and spans a wide metallicity interval, from  $\Zini=0.0005$ to $\Zini=0.03$ (typically 10 values). Therefore, each set includes roughly 700 TP-AGB tracks, computed up to the complete ejection of the envelope.
For the purposes of this work the grid of evolutionary TP-AGB models consists of 33 sets ($\sim$ 23000 TP-AGB tracks). Such a demanding  computational effort can only be achieved with a flexible and fast code such as \colibri.
The full grid of TP-AGB tracks is summarised in Table~\ref{tab:mod_grid}.

\begin{table*}
\caption{Grid of TP-AGB sets}\label{tab:mod_grid}
\input{tpagb_grid}
\end{table*} 

The observed and simulated number counts for each class of AGB stars, the corresponding \chisqlf\ and the C/M ratio, i.e. the number of C-type (C-rich and X-AGB) to M-type (O-rich) AGB stars, are given in Table~\ref{tab:res_CM}.
All the LFs calculated for each set are presented in Appendix C available online.
In Fig.~\ref{fig:res_2} we show a summary of the performance of the calculated models and we illustrate the fundamental steps of the calibration in the next section.

\begin{table*}
\begin{minipage}{170mm} 
\caption{Comparison of observed star counts with \colibri\ models.} 
\label{tab:res_CM}
\input{SMC_dpmod_amc15sic_tab_results}
\end{minipage}
\end{table*}

\begin{figure*} 
\centering
\includegraphics[width=0.7\textwidth]{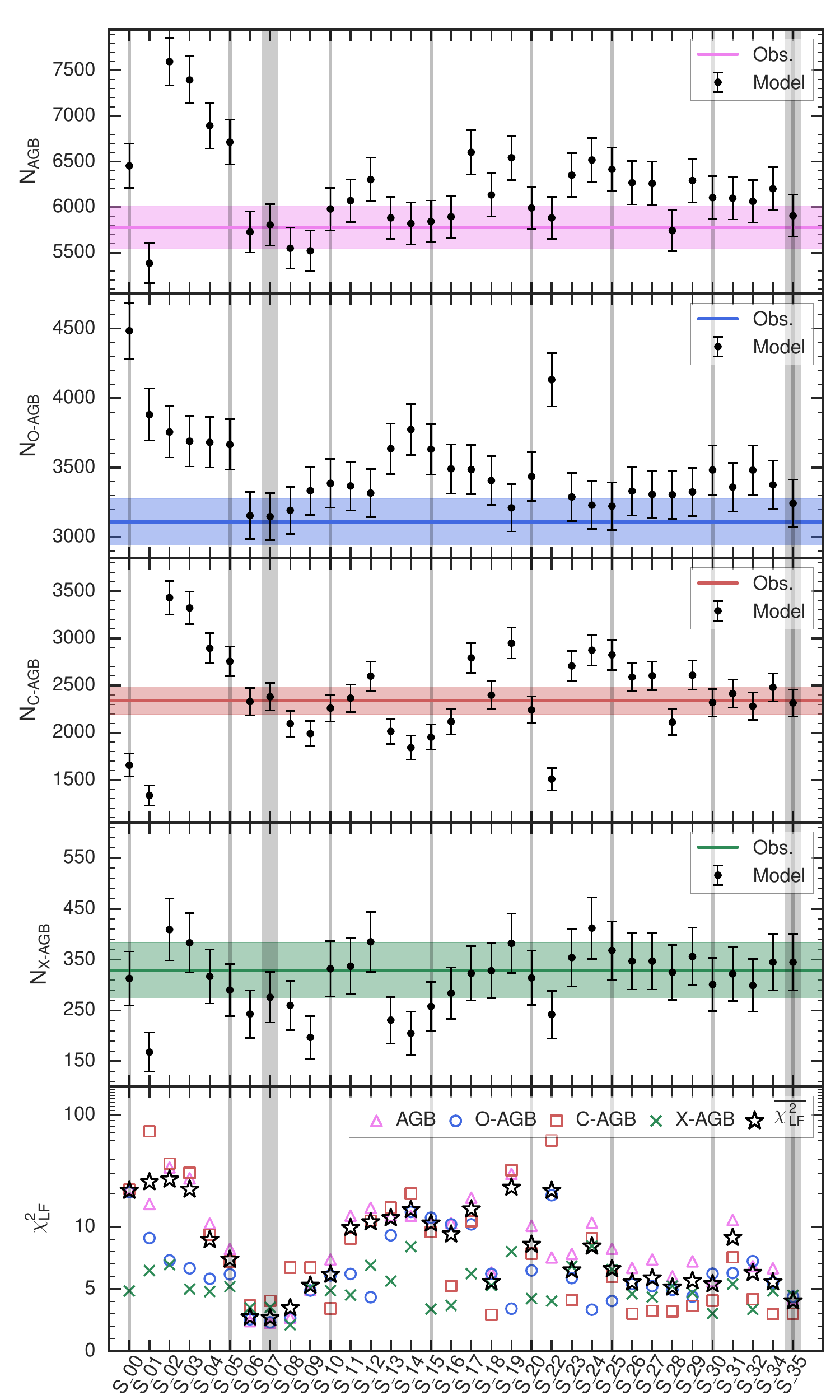}
 \caption{Summary of the results obtained from the calculated models. In the first four panels, we show the number of simulated stars and the 3$\sigma$ error bars for each set of models. The horizontal lines represent the number of observed stars and the relative 3$\sigma$ error bars (estimated as the square root of the number counts). The bottom panel shows the resulting \chisqlf\ values for the AGB, O-, C- and X-AGB samples. The average value $\overline{\chi^{\,2}_{\rm \,LF}}$ of the four \chisqlf\ for each set is shown with the star symbols. The two well defied minima, highlighted with the vertical grey strips, correspond to the two best-fitting models (S\_07 and S\_35). The vertical thin grey lines marked sets S\_00, S\_05, S\_10, S\_15, S\_20, S\_25, S\_30 and S\_35.}
\label{fig:res_2}
\end{figure*}

To facilitate the presentation of the calibration it is convenient to split the TP-AGB sets into two groups, depending on the assumptions adopted for the two key processes under examination, namely mass loss (see Sect.~\ref{ssec:changemdot}) and 3DU (see Sect.~\ref{ssec:changedup}).

\subsection{Changing mass-loss}
\label{ssec:changemdot}

The first series of models  (S\_01 -- S\_07) shares the same prescriptions for the 3DU. In particular, the efficiency $\lambda$ follows the formalism introduced by K02, which assigns quite high values ($\lambda \simeq 1$) to stars with initial masses larger than $\simeq 3$~\Msun.

The initial step is to reduce the excess of O-rich stars present in our starting set S\_00. 
This goal requires increasing the efficiency of mass loss both during the pre-dust and the dust-driven regimes.
This is partly achieved with the adoption of the CS11 formalism to describe the cool chromospheric mass loss during the early stages when dust is expected not to be the main driver of stellar winds. 
In fact, the pre-dust mass loss is already active prior to the 1TP, i.e. during the brightest part of the early-AGB. For instance, in the case of a 1~\Msun\ model with $\Zini=0.001$ the current mass at the beginning of the TP-AGB phase is reduced by $\approx 15 $ per cent. The net effect is the shortening of the duration of the TP-AGB phase, mostly for low-mass O-rich stars, thus reducing their numbers. 
All S\_01 -- S\_07 sets exhibit a reduction in the number of O-rich stars compared to the set S\_00, improving the  comparison with observations.

To explore the impact of mass loss in the dust-driven regime, the set S\_01 assumes a very efficient mass loss, adopting BL95 with $\etadust=0.05$ for all models. However, the reduction
of O-rich stars is accompanied by a further deficit in the predicted number of C-rich stars, which worsens the comparison with observations compared to S\_00.

In all sets from S\_02 -- S\_07  we vary the mass-loss laws in the dust-driven regime, depending on the surface C/O ratio: for C/O~$< 1$ we use the relation proposed by BL95, while for C/O~$>1$ we use a routine based on the recent results of dynamical atmosphere models \citep[][hereinafter CDYN]{Eriksson_etal14, mattsson10}.
All these models adopt the same CDYN prescription for \mdotdust\ when C/O~$> 1$, while the efficiency of the BL95 relation, active for C/O~$< 1$, is increased from $\etadust=0.02$ (for S\_02) to $\etadust=0.06$  (for S\_07). 
Interestingly, this variation results in a relatively modest effect on the total number of O-rich stars, but has a significant impact on the predicted C-rich LF. This effect can be appreciated in Fig.~\ref{fig:res_2} which shows that the total number of O-rich stars do not vary significantly along the sequence from S\_02 to S\_05, whereas the total number of C-rich (and X-AGB) stars progressively decreases (see also Figs. C.3, C.4, C.5 and C.6 in the online Appendix C).

The reason of the small effect on the number of O-rich stars is that the bulk of them have low-mass progenitors, i.e. $\Mini \lesssim 2$~\Msun, for which the duration of the AGB phase is mainly controlled  by the cool chromospheric mass loss active  before the TP-AGB phase.
Due to its high luminosity dependence, the BL95 relation mainly affects the evolution of more massive (and brighter) AGB stars that experience HBB. These represent a small contribution to the total number of O-rich stars (see Sect.~\ref{sec:HBBimpact} below).

On the other hand, the sizeable impact on the number of C-rich stars can be explained as follows.
The activation of the CDYN \mdotdust\ takes place once specific physical thresholds are met, i.e. sufficiently low effective temperatures, enough carbon excess to be condensed into dust grains, and suitably large L/M ratios. In general, increasing the efficiency of mass loss during the O-rich phase tends to favour larger mass-loss rates also during the stages with C/O~$> 1$, since TP-AGB stars will enter the C-rich phase with  lower current masses and lower effective temperatures.  

However, we should also note that extremely efficient mass loss during the O-rich stages may even inhibit the formation of C-stars by either anticipating the termination of the TP-AGB phase, or preventing the star from attaining the minimum temperature, \Tbdred\, for the onset of the 3DU if the envelope mass has been significantly reduced.
Indeed, this circumstance is met in the evolutionary models of the set S\_07, so that the maximum initial mass for the formation of C-stars at $\Zini = 0.004$ is found to be around 2.6~\Msun\ (see Sect. \ref{ssec:mini_cstars}). 

\begin{figure}
\centering
\includegraphics[width=0.9\columnwidth]{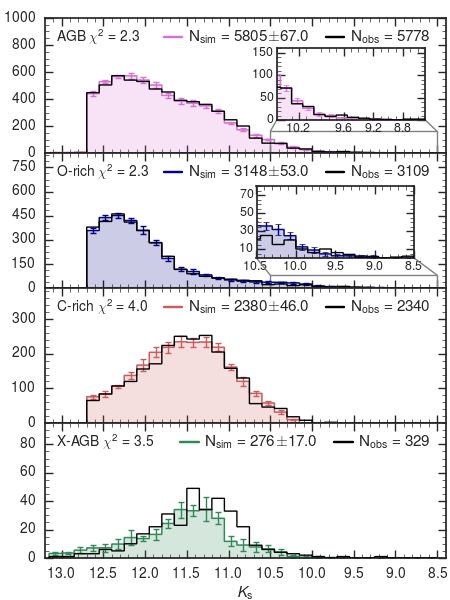}
\caption{Same as Fig.~\ref{fig:LFKsS000}, but for the best-fitting set S\_07.}
\label{fig:LFKsS007}
\end{figure}

All LFs computed with an efficiency $\etapre=2$ (as in the sets S\_02 -- S\_05)  show an excess ($\approx 20$ per cent) of low-mass O-rich stars that populate the faintest bins ($\ks \gtrsim 11.8$~mag).
This leads us to increase the \etapre\ in the CS11 prescription, passing from  $\etapre=2$ to  $\etapre=3$, as in sets S\_06 and S\_07. As can be appreciated in Table~\ref{tab:res_CM} and in Fig.~\ref{fig:res_2}, this modification is successful and the residual excess of faint O-rich stars is removed. 
Models S\_06 and S\_07 only differ in the choice of the \Tbdred\ parameter, which increases linearly with the initial $\Zini$ in the case of S\_07 (see notes of Table~\ref{tab:mod_grid}). 
This assumption is equivalent to reducing the formation of C-stars at higher metallicity, as indicated by observations of the C/M ratio in galaxies  \citep[e.g.][]{boyer13, CioniHabing_03, Groenewegen_02}, 
and by  earlier population synthesis studies \citep{marigo07, Marigo_etal99}. Anyhow, we note that at SMC metallicities the  
differences in the \Tbdred\ parameter are negligible, becoming significant only at lower and higher $\Zini$.
 
In addition to the BL95 relation for mass loss, we run a set of models (S\_19) in which we adopt the formalism of VW93 for the efficiency of dust-driven winds during the O-rich stages, while keeping the CDYN prescription for C-stars.
The LF of O-rich stars is recovered very well, with a relative difference in the number counts within $\sim 3$ per cent. However, the same model leads to an excess of C-stars (in the brighter bins, $\ks \la 10.5$ mag) of about 25 per cent. 

Among all the sets of this first series we identify the best-fitting model as the set S\_07, for which we obtain the lowest \chisqlf\ values. The resulting LFs are shown in Fig.~\ref{fig:LFKsS007}. This set recovers quite well both the LFs of all classes of AGB stars and the number counts. 
The optimal combination of parameters includes: CS11 pre-dust 
mass loss with efficiency $\etapre=3$,  dusty regime with BL95 and $\etadust = 0.06$ for $\co\leq1$ and the CDYN for $\co>1$; 3DU efficiency described following K02  (see Table~\ref{tab:mod_grid}).

\subsection{Changing the third dredge-up}
\label{ssec:changedup}

 \begin{figure*}
 \includegraphics[width=\textwidth]{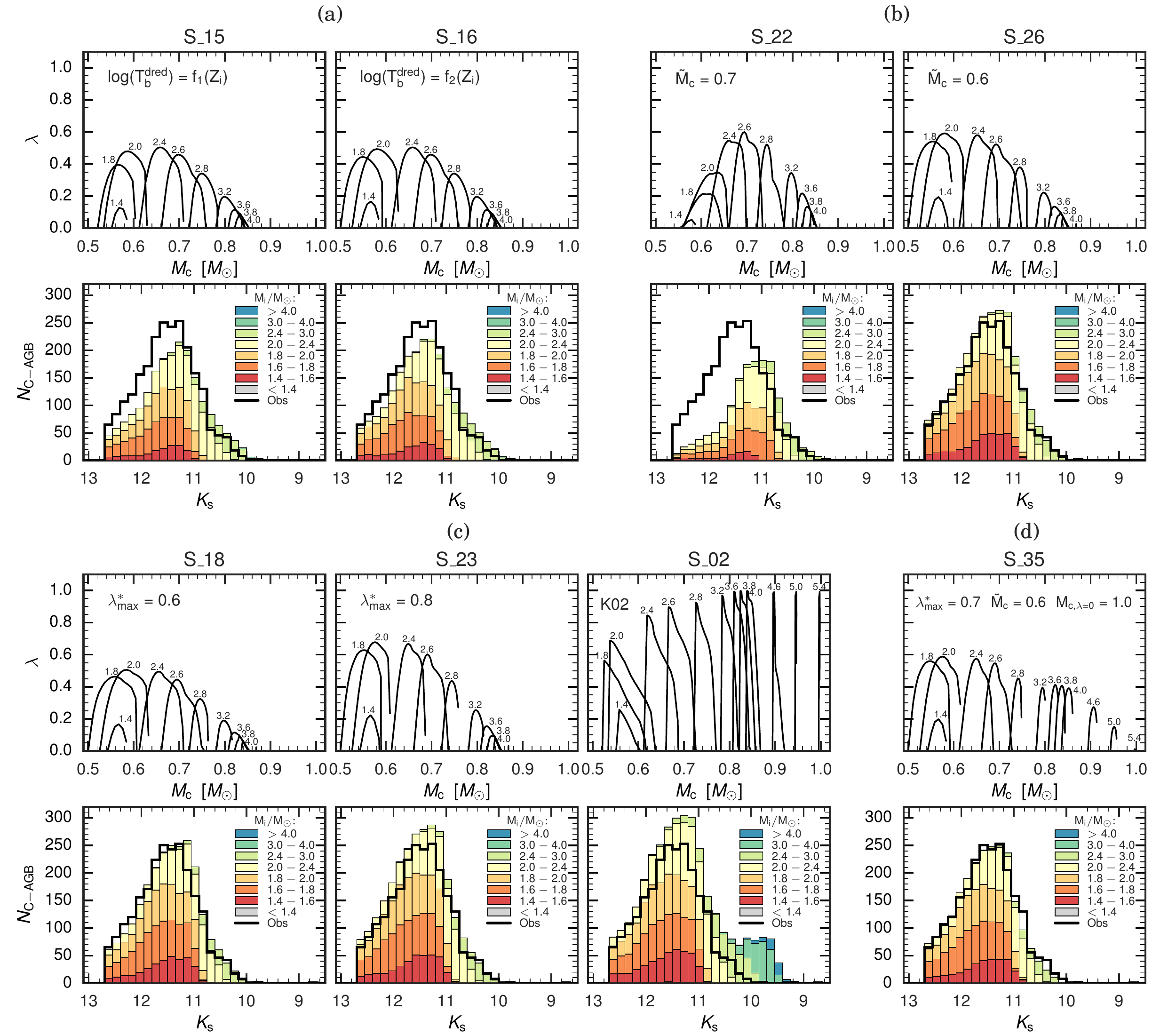}
    \caption{Top rows of each panel: efficiency of the 3DU ($\lambda$) as a function of the core mass $M_\mathrm{c}$ of a few selected evolutionary tracks with $\Zini=0.004$ and initial mass as labelled in the figure. Bottom rows of each panel: observed (black histograms) and simulated CSLFs as derived from the corresponding above sets of models. The synthetic LFs are shown as stacked histograms to highlight the contribution of each initial mass bin to the LF as indicated in the legend. Panel (a): these models (S\_15 and S\_16) share the same input prescriptions but for the temperature parameter (as indicated) that controls the onset of the 3DU. Panel (b): models S\_22 and S\_26, which differ by the core mass, $\widetilde{M}_\mathrm{c}$, that corresponds to the maximum 3DU efficiency. Panel (c): moving rightward the former two sets (S\_18 and S\_23) differ in  $\lambda_{\rm max}^{\ast}$, while the latter two sets (S\_23 an S\_02)  mainly differ by the efficiency of the 3DU in TP-AGB stars of higher initial mass, i.e. $\Mini \ga 3 \, \Msun$. Panel (d): best-fitting model of this series (S\_35) which has the same input prescriptions of S\_26 but for the  $M_{\mathrm{c},\lambda=0}$ and the BL95 efficiency ($\eta = 0.03$).} 
    \label{fig:lambda_plot_1fig}   
\end{figure*}

The first part of our work shows that, adopting our standard prescriptions for the 3DU, based on \citet{karakas02} models, the AGB star population in the SMC can be well reproduced with suitable choices for the mass-loss regimes. 
However, at this stage we cannot consider the calibration work complete. Given the lack of a robust treatment of stellar convection and mixing in stellar evolution codes it is not sufficient to limit our investigation to a single 3DU prescription. 

As already mentioned in Sect.~\ref{ssec:3dup} and shown in Fig.~\ref{fig:fruity_lamb}, the predicted properties of the 3DU are quite heterogeneous in the literature. The largest differences show up at larger stellar masses, typically for $\Mini \ga 3$~\Msun.
In this mass range the  K02 formalism predicts a very efficient 3DU, $\lambda \simeq 1$, for all models with $M > 3$~\Msun, irrespective of metallicity. TP-AGB evolutionary calculations by other authors
predict lower efficiencies of the 3DU, with typical values $\lambda \la 0.5$  for $M_{\rm c} > 0.8$~\Msun\ \citep{Cristallo_etal15,VenturaDantona_09}.

We should note that hints for a reduced efficiency of the 3DU in intermediate-mass AGB stars ($\lambda < 1$ for $\Mini > 3-4\, M_{\odot}$) come also from studies on the initial-to-final mass relation (IFMR) derived from white dwarfs (WD) data in Galactic open clusters (see Fig.~\ref{fig:ifmr}). As already pointed out by \citet{Kalirai_etal_14}, current TP-AGB models tend to predict WD masses that are too low for initial masses $\ga 3$~\Msun. A significant improvement is obtained by reducing the efficiency of the third dredge-up during the TP-AGB phase of progenitor stars.
In fact the 3DU has a direct impact on the evolution of the  core mass reducing its net growth \citep[e.g.][]{Kalirai_etal_14, marigo13, Herwig_04}. Therefore, lowering the 3DU in intermediate-mass stars may help to obtain more massive WDs.

In the following  we will focus our discussion on those sets that share the same description for the dust-driven mass-loss regime. This relies on two alternatives depending on the C/O ratio, namely: the BL95 relation for O-rich stars and the CDYN option for C-rich stars.
This latter represents our preferred option as it is based on well-tested and state-of-the-art dynamical atmosphere models for C-stars \citep{Eriksson_etal14, mattsson10}.

Given these premises, we now analyse the impact of varying the 3DU description in our simulations.
As an initial step we explore the effect of keeping the K02 relations  for $\lambda$ while 
lowering \Lambdamax\  down to a fixed value of 0.4-0.5 (S\_08, S\_11, S\_31), i.e. we impose that the 3DU efficiency does not exceed the selected \Lambdamax in all models. 
By analysing the C-rich and X-AGB LFs we find that, unless we assume a very efficient mass loss ($\etadust=0.06$ as in S\_08) during the O-rich dusty regime, the models predict an excess of bright C-rich stars ($\ks \lesssim 10$~mag), most evident in S\_11 and S\_31.

To carry out a more systematic exploration on the effects of varying the 3DU efficiency, 
we relax our standard choice, based on K02, and adopt the new formalism, introduced in Sect.~\ref{ssec:3dup} (sets from S\_13 to S\_35). Its parametric formulation enables us to investigate various choices of the 3DU law, including  also the possibility that the efficiency of the mixing episodes become lower at larger stellar masses.
Since we have already explored the cases of extremely efficient mass loss for 
O-rich stars (first series of models in Sect.~\ref{ssec:changemdot}), we now opt to
reduce \etadust\ down to 0.01-0.03, values that are also adopted in other widely used AGB models \citep[e.g.][]{karakas_18, VenturaDantona_09}.

In Fig.~\ref{fig:lambda_plot_1fig} we show the resulting carbon star luminosity functions (CSLFs) for a few sets, together with the evolution of $\lambda$ as a function of the core mass $M_\mathrm{c}$ for some selected TP-AGB evolutionary tracks at $\Zini=0.004$, representative of the initial metallicity for the bulk of C-rich stars. 
The emerging picture is quite complex due to the interplay of the 3DU with mass loss, still we can extract a few key general indications that help the interpretation of the results.

\begin{itemize}
\item [(i)] \textit{The faint tail} 
contains information about the lowest-mass stars that become C-stars.  This limit is controlled by the temperature parameter $\Tbdred$, as well as by $\widetilde{M}_\mathrm{c}$. As illustrated in panel (a) of Fig.~\ref{fig:lambda_plot_1fig}, lowering $\Tbdred$, i.e. moving from set S\_15 to set S\_16, favours an earlier formation (i.e.\ at lower $M_\mathrm{c}$ and $L$)  of C-stars with low masses and therefore increases the predicted number of C-stars in the faintest bins of the CSLF. Decreasing $\widetilde{M}_\mathrm{c}$ has a similar effect on the faint tail, as shown in panel (b) of Fig.~\ref{fig:lambda_plot_1fig}.\\
 
\item [(ii)] \textit{The position of the peak} is mainly affected by the choice of $\widetilde{M}_\mathrm{c}$ as shown in panel (b) of Fig.~\ref{fig:lambda_plot_1fig}. At lower $\widetilde{M}_\mathrm{c}$, i.e.\ moving from set S\_22 to set S\_26, the CSLF peak shifts towards fainter magnitudes. At the same time, lowering the value of $\widetilde{M}_\mathrm{c}$ results in an increase of the total number of C-stars, in particular those that populate the faint tail of the CSLF.\\

\item [(iii)] \textit{The amplitude of the peak} is mainly controlled by the choice of $\lambda_{\rm max}^{\ast}$.  As shown in panel (c) of Fig.~\ref{fig:lambda_plot_1fig}  increasing the maximum 3DU efficiency, i.e. moving from set S\_18 to set S\_23,  produces a higher peak in the CSLF.\\

\item [(iv)] \textit{The bright end} contains information about the highest-mass stars that become C-stars, and is controlled by $M_{\mathrm{c},\lambda=0}$ together with the efficiency of mass loss. As illustrated in panel (c) of Fig.~\ref{fig:lambda_plot_1fig}, increasing the value of $M_{\mathrm{c},\lambda=0}$ , i.e.\ moving from set S\_23 to set S\_02, causes the appearance of C-stars at brighter magnitudes. It 
essentially reflects the larger $\lambda$ attained by stars with $\Mini \ga 2.5$~\Msun in the present models. 

We may conclude that with a very 
efficient 3DU in more massive AGB stars ($\lambda\approx 1$, as in K02) models 
tend to largely over-predict the number of bright C-stars (rightmost plot of Fig.\ref{fig:lambda_plot_1fig}, panel (c), for the set S\_02).
As already discussed in Sect.~\ref{ssec:changemdot}, such excess can be removed only by invoking a powerful  mass loss during the preceding O-rich stages (e.g., as for S\_07 in the first series of sets).
\end{itemize}

\begin{figure}
\centering
\includegraphics[width=0.9\columnwidth]{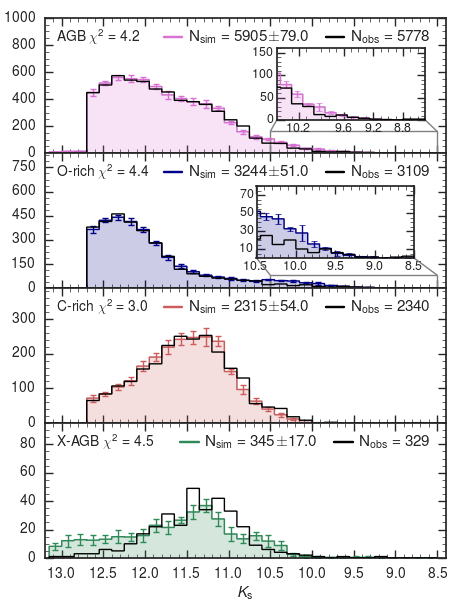}
\caption{Same as Fig.~\ref{fig:LFKsS000}, but for the best-fitting set S\_35.}
\label{fig:LFKsS035}
\end{figure}

Among all sets of the second series, the best-fit is provided by S\_35 which is found to reproduce well all LFs (see Fig.~\ref{fig:LFKsS035}) and the corresponding number counts within 5 per cent. The optimal combination of parameters includes: CS11 pre-dust 
mass loss with efficiency $\etapre=3$,  dusty regime with BL95 and $\etadust = 0.03$ for $\co\leq1$ and the CDYN for $\co>1$; $\lambda_{\rm{max}}^{\ast} = 0.7$, $\widetilde{M}_\mathrm{c}$ = 0.6, (see Table~\ref{tab:mod_grid} and panel (d) of Fig.~\ref{fig:lambda_plot_1fig}).
 
In the next section we discuss the results of the two best-fitting models S\_07 and S\_35 and the main differences with respect to the starting set S\_00.

\section{The best-fitting models}
\label{sec:bestfit}

Our final results are summarised in Table~\ref{tab:res_CM} and Fig.~\ref{fig:res_2}. Models S\_07 and S\_35 stand out clearly by their smaller \chisqlf\ and indeed provide an excellent description of the observed CMDs (see Fig.~\ref{fig:obs_S35_co}) and LFs (see Figs.~\ref{fig:LFKsS007} and 
\ref{fig:LFKsS035}). In the following we discuss the main implications we may derive from the analysis of the results.
The main differences between the two sets deal with the chemical composition of the ejecta and the final core masses. While  a proper analysis on the predicted chemical yields is postponed to a follow-up work (Marigo et al., in prep.), we briefly discuss the IFMR in the next Sect.~\ref{ssec:ifmr}.

\subsection{The IFMR relation}
\label{ssec:ifmr}

\begin{figure}
\centering
\includegraphics[width=\columnwidth]{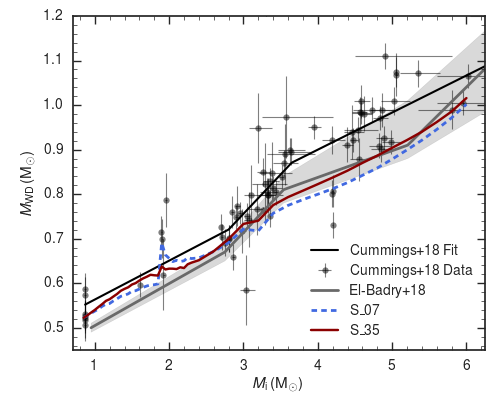}
\caption{Initial-final mass relation for white dwarfs in the solar neighbourhood.
The grey line and the shaded region show the best-fitting IFMR and its 95 per cent probability, respectively,  derived from \textit{Gaia} data \citep{gaia_ifmr}. The \parsec\ based semi-empirical data
and the 3-pieces fit derived by \citet{cummings_18} are shown as grey
points with relative error bars and a solid black line, respectively. The IFMRs
for $\Zini = 0.014$ (the solar metallicity) derived from S\_07 and S\_35 are shown as blue dashed and red solid lines, respectively.}
\label{fig:ifmr}
\end{figure}

Two recent semi-empirical IFMR together with model predictions from the two best-fitting sets are shown in Fig.~\ref{fig:ifmr}. 

While the general trend is satisfactory for both sets S\_07 and S\_35, at larger initial masses, $M > 3 M_{\odot}$, the S\_07 models are only marginally consistent with the semi-empirical IFMRs. 
The WD masses for intermediate-mass stars ($\Mini  >  3\, \Msun$) appear to be underestimated by the models, in comparison with both relations presented by \citet[][]{cummings_18} and \citet[][]{gaia_ifmr}.
The predicted IFMR based on the set S\_35 improves the comparison, mostly with respect to the initial masses larger than about 3~\Msun . The larger WD masses predicted by these models, compared to those of the set S\_07, result from reducing the efficiency of both the third dredge-up  and mass loss during the  TP-AGB phase of the progenitors. 
We also note that the agreement becomes quite good with the semi-empirical relation based on \textit{Gaia} DR2 
data \citep{gaia_ifmr}, whereas a substantial discrepancy still affects the comparison with the IFMR
from \citet{cummings_18}. This latter runs systematically above the mean relation from \textit{Gaia} data. 
The reason for such a large difference between the two semi-empirical relations is not known at present. More work is required to clear up this point, given its importance for the calibration of the AGB stellar evolution models.

\subsection{The initial masses of C-stars}
\label{ssec:mini_cstars}

\begin{figure*}
\centering
\includegraphics[width=\textwidth]{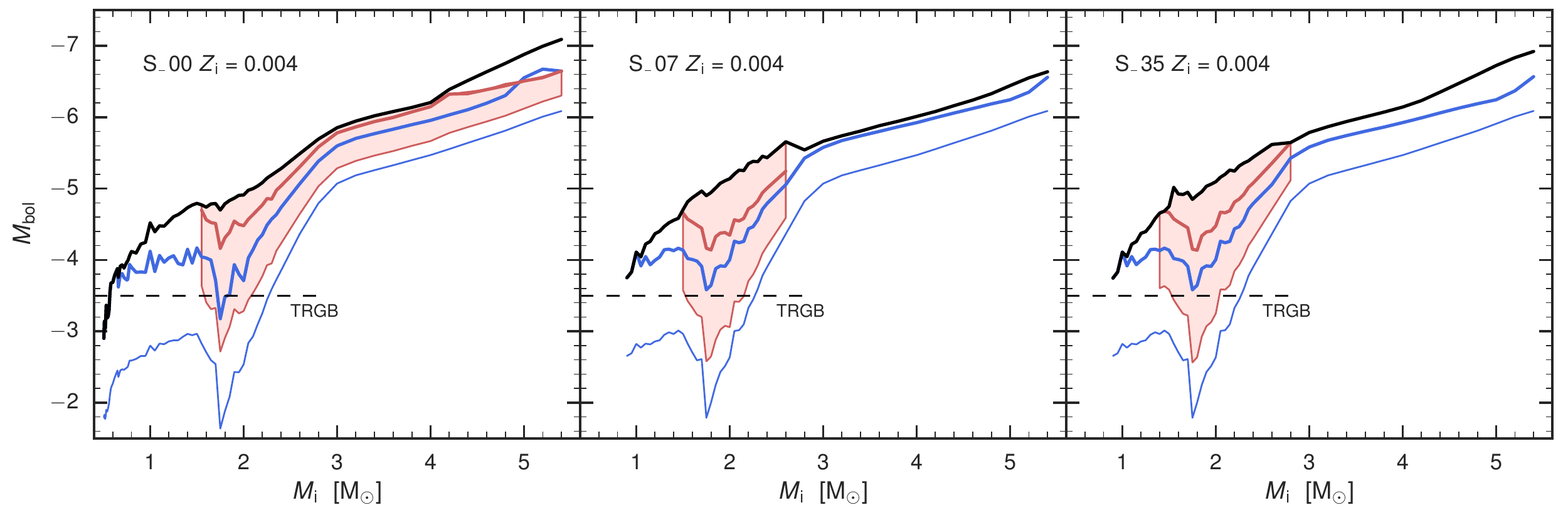}
\caption{Bolometric magnitudes as a function of \Mini\ for a few relevant transition stages: the first TP (blue), the transition from the O-rich to the C-rich domain (red), and the AGB tip (black). In the first two cases,
thick solid lines correspond to the luminosities at the quiescent stages that precede TPs, while thin solid lines correspond to the faintest luminosities reached during the post-TP low-luminosity dips. 
Results are shown for the TP-AGB sets S\_00, S\_07, and S\_35 with $\Zini=0.004$. }
\label{fig:mbol_tr_lum}
\end{figure*}

Figure~\ref{fig:mbol_tr_lum} shows the predicted ranges of initial masses and bolometric magnitudes of C-stars for the starting set S\_00 and the two best fitting sets S\_07 and S\_35.

First, it is striking that in both S\_07 and S\_35 models, C-stars are expected to form only in a limited interval of initial masses, between 1.4 and 2.8~\Msun, at SMC-like metallicities (i.e. $\Zini =0.004$).
In particular,  the upper limit for C-stars formation at around 3~\Msun\ is mainly constrained  by the observed deficit of C-rich stars for magnitudes brighter than $\ks \simeq 10$~mag.  Our calibration indicates that C-stars with  initial masses $\gtrsim 3$~\Msun\ would form a bright red tail that is not actually observed in the 2MASS CMD.

The two best-fitting sets converge to the same threshold mass following somewhat different paths. In the case of S\_07, the formation of massive C-stars is prevented because of a high mass-loss efficiency that drastically shorten the lifetimes, despite the efficient 3DU based on K02 formalism. 
In the case of S\_35, the reduced 3DU efficiency -- assumed for TP-AGB models of higher masses -- contributes to confining the formation region of C-stars to lower initial masses.
In both sets of models the upper mass limit for C-stars formation is controlled by the efficiency of mass loss and 3DU, rather than by the onset of HBB.

The results of our calibration need to be compared with the predictions available in the literature.
Past works that try to reproduce the CSLF in the SMC present much wider ranges of masses for the formation of C-stars. For instance, \citet{groenewegen93} and \citet{marigo07} find C-stars still forming at initial masses as high as 4~\Msun. 
Also, recent AGB models, more focused on the chemical yields, predict that the maximum initial mass for C-star formation at SMC-like metallicity lies between between 4.0 and 4.5~\Msun\ \citep[][$\Zini=0.028$]{Karakas_etal18}, or up to 6~\Msun\ \citep[][$\Zini=0.003$]{Cristallo_etal15}. On the other hand, AGB models presented by \citet{VenturaDantona_09} and \citet{dellagli15smc} form carbon stars with initial masses up to $\simeq 3\, \Msun$ at $\Zini=0.004$, a value that is in close agreement with our calibration.

It is also interesting to compare these predictions with the cluster data in the Magellanic Clouds. According to the classical compilation by \citet{frogel90}, `the youngest clusters in which C-stars are found have an age of about 100 Myr implying a maximum initial mass for these stars of 3--5~\Msun'. This conclusion appears in contradiction with our results; however, it was based on quite crude age estimates for the clusters, and uncertain memberships for the C-stars. According to the revised compilation by \citet{girardi07}, the youngest LMC cluster to contain a C-star in its central region 
is NGC~1850, which has isochrone ages of  $\sim\!60$--$90$ Myr \citep{correnti17} and hence turn-off masses $\gtrsim5.5$~\Msun. This high turn-off mass appears compatible with AGB stars that experience
efficient HBB, a process that usually prevents the formation of C-stars; therefore the C-star in NGC~1850 is usually regarded as a chance alignment of a field LMC star. The second youngest LMC clusters to contain C-stars are NGC~1987 and NGC~2209, with one and two C-stars, respectively. NGC~1987 has estimated ages between $\sim\!1$ \citep{goudfrooij17} and 1.3--1.5 Gyr \citep{milone09}, while NGC~2209 has ages of $\sim\!1.15$~Gyr \citep{correnti14}. Both clusters therefore have turn-off masses slightly below 2~\Msun. 
We remark that NGC~2209 was regarded as a much younger cluster (ages between 120 and 370 Myr, hence turn-off masses between 4 and 2.7~\Msun) in the work by \citet{frogel90}. In the SMC, the revised cluster ages by \citet{glatt08} indicate that the youngest such cluster is the $\sim\!1.5$-Gyr old NGC~419, with its impressive population of ten C-stars. The presence of a double red clump in this cluster \citep{girardi09} firmly points to turn-off masses of $\sim\!1.75$~\Msun\
\citep{girardi13}. 
Overall, the revised cluster data do not appear to indicate the presence of C-stars with progenitors more massive than about 2~\Msun. On the other hand,
Magellanic Clouds star clusters with turn-off masses between 2 and 3~\Msun\ are rare and more scarcely populated than older clusters, so that the maximum turn-off mass giving origin to C-stars cannot be identified with confidence from these data alone. 

Fig.~\ref{fig:mbol_tr_lum} shows other aspects that are worth note.
The faintest transition to the C-star domain happens at $\Mini \simeq 1.8~\Msun$. This value coincides (at $\Zini =0.004$)
with the initial mass boundary between stars that develop degenerate
He-cores after the main sequences and the more massive ones
that skip electron degeneracy.
As discussed by \citet[][]{Lebzelter_etal18},
the presence of a well-defined minimum in the M-to-C transition luminosity
at increasing stellar mass explains the splitting of the data
along two branches \citep[labelled (a) and (c) in fig.~3 of][]{Lebzelter_etal18}
in the diagram that combines 2MASS and \textit{Gaia} photometry for long period variables.

We expect that some fraction of TP-AGB stars  may appear below the RGB-tip (at $\Mbol \approx -3.5$ mag), mostly during the relatively long-lived low-luminosity dip that follows the occurrence of a thermal pulse \citep[see e.g.,][]{BoothroydSackmann_88}.
These faint stars may be either O-rich or C-rich.
The O-rich stars below the RGB-tip span a range of initial masses $1.0~\Msun \la \Mini \la 2.2~\Msun$ and cover a fraction of $\approx$ 26-27 per cent of the entire O-rich TP-AGB population, for both sets S\_07 and S\_35.
The C-rich stars below the RGB-tip should correspond to initial masses  in the range 
$1.5~\Msun \la \Mini \la 2.0~\Msun$. Our simulations indicate these faint C-stars represent $\approx$  7 per cent of the entire C-rich population, for both  sets S\_07 and S\_35.  

\subsection{Lifetimes}
\label{ssec:lifetimes}

\begin{figure*}
\centering
\includegraphics[width=\textwidth]{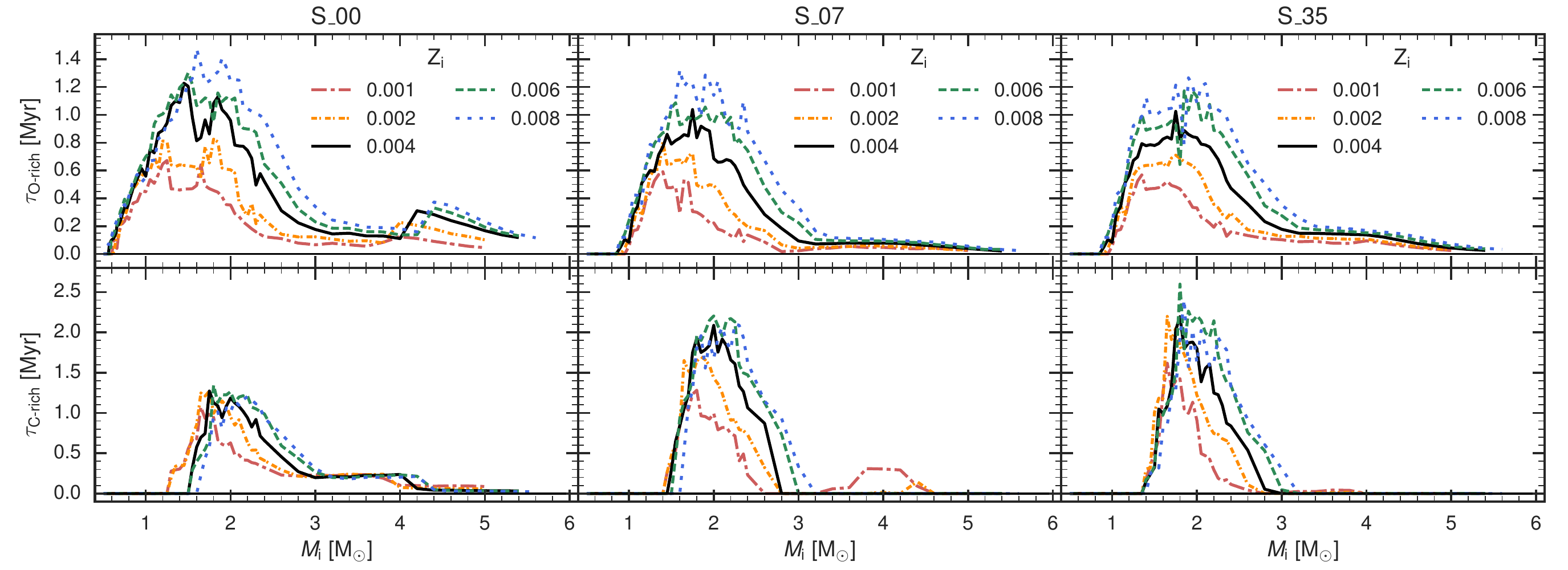}
\caption{TP-AGB lifetimes of O- and C-rich stars for selected values of initial metallicities as predicted by the initial set S\_00 (left panels) and the two best-fitting sets S\_07 (middle panels) and S\_35 (right panels). }
\label{fig:lifetimes}
\end{figure*}

TP-AGB lifetimes are relevant to quantify  the contribution of low- and  intermediate-mass stars to i) the chemical enrichment of the interstellar medium, and ii) the integrated light of galaxies \citep{Marigo_15}. In fact, the duration of the TP-AGB phase, $\tau_{\rm TP-AGB}$, determines the level of chemical enrichment of the ejecta, which is set through the number of thermal pulses and 3DU episodes experienced by a star and/or the period in which HBB is operating.
At the same time   $\tau_{\rm TP-AGB}$
controls the energy emitted during the phase through the relation $E_{\rm TP-AGB} = \int_{\tau_\mathrm{TP-AGB}} L(t) dt$, where $L(t)$ is the stellar luminosity at time $t$.  

In Fig.~\ref{fig:lifetimes}, we present the comparison between the predicted lifetimes of M- and C-stars as derived from the starting set of TP-AGB models S\_00 and the two best-fitting models S\_07 and S\_35. We consider only models with luminosity higher than $\log (L/L_{\odot}) = 3.3$, i.e. brighter than the RGB tip.

As in other models in the literature \citep[e.g.][]{weiss_ferguson_09, Karakas_10}, the predicted TP-AGB lifetimes peak at $\Mini \simeq 2 $~\Msun, close to the value below which stars develop a degenerate helium core after the main sequence, the exact value depending on the initial metallicity and other model details. 
The general trends of the lifetimes for O-rich stars are similar in the three set of models, yet some differences are present for $\Mini \gtrsim 3$~\Msun. Compared to set S\_00, the best-fitting models S\_07 and S\_35 predict shorter lifetimes by a factor of $\approx 2$.
As to C-stars, we note an increase of the lifetimes in the new models
compared to set S\_00, reaching a factor of 2 near the peak at $\Mini \sim 2$~\Msun.

Within the metallicity range covered by the calibration ($0.001 \la \Zini \la 0.004$), we expect shorter O-rich stages at decreasing  $\Zini$. This is the consequence of the earlier transition to the C-rich regime. 
The lifetimes for C-rich stars also follow a decreasing trend with lower $\Zini$, particularly evident for initial masses close to the peak and larger than $\approx 2$~\Msun. The reason is linked to the fact that, at a given initial mass, a TP-AGB star of lower metallicity is characterized by a larger core mass. As a consequence,  the TP-AGB phase proceeds at higher luminosities and with shorter inter-pulse periods, i.e. thermal pulses and 3DU episodes take place more frequently with consequent quicker increase of the carbon excess during the C-star stages. Both factors (related to luminosity and carbon excess) tend to favour an earlier onset of the dust-driven wind in C-stars of lower metallicity, in the range under consideration.

\subsection{The overall impact of HBB}
\label{sec:HBBimpact}
 
\begin{figure}
    \centering
    \includegraphics[width=\columnwidth]{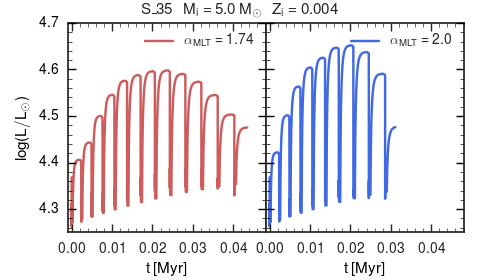}
    \caption{Luminosity as a function of time for TP-AGB tracks of 5~\Msun\, and $\Zini = 0.004$ calculated using the input parameters of set S\_35 with the solar-calibrated $\alpha_\mathrm{MLT}=1.74$ (left panel) and $\alpha_\mathrm{MLT}=2.00$ (right panel). Time is set to zero at the first thermal pulse.}
    \label{fig:alpha_5Msun}
\end{figure}

As already mentioned, our models assume a constant and solar-calibrated mixing-length parameter, with $\alpha_\mathrm{MLT}=1.74$, which turns out to affect the efficiency of HBB in the models. Since there are significant claims that such a convection treatment could be inaccurate and that $\alpha_\mathrm{MLT}$ (or the equivalent parameter coming from different approaches) may vary along the sequences of red giants \citep[e.g.][and references therein]{tayar17}, let us verify the impact of such an assumption in our results.

First of all, it turns out that adopting $\alpha_\mathrm{MLT}=1.74$ in our best-fitting S\_07 and S\_35 sets, stars with initial masses larger than 3~\Msun\, represent a little contribution to the entire AGB star population of the SMC, $\simeq 8$ and $\simeq 9$ per cent respectively. 
Therefore we expect that, overall, stars in this mass range should have a  small impact in our calibration.

To test this point in detail we run an additional set of AGB models over the relevant range of metallicity, from $\Zini = 0.001$ to $\Zini = 0.008$,  keeping the same input prescriptions as in the set S\_35, except for the  mixing-length parameter which is changed from $\alpha_\mathrm{MLT}=1.74$ to $\alpha_\mathrm{MLT}=2.00$ in the models with initial masses $> 3~\Msun$.
The main effects of increasing $\alpha_\mathrm{MLT}$ are those of i) lowering the minimum stellar mass for the activation of  HBB  and ii)  making stars with HBB hotter and more luminous, as can be seen in Fig.~\ref{fig:alpha_5Msun}. Given the significant luminosity dependence of the B95 mass-loss formula, models with  $\alpha_\mathrm{MLT}=2.00$ have a shorter duration of the AGB phase. At $\Zini =0.004$ the reduction of the TP-AGB lifetimes does not exceed $\simeq 30$ per cent in most cases. The impact on  star counts is relatively modest, so that AGB stars with $\Mini > 3~\Msun$  make an overall contribution of $\simeq 6$ with $\alpha_\mathrm{MLT}=2.00$,  in place of $\simeq 9$ predicted with $\alpha_\mathrm{MLT}=1.74$.
Therefore, this test confirms that uncertainties in the efficiency of HBB, due to reasonable changes of the mixing-length parameter, do not affect the results of the present calibration.

\subsection{Integrated luminosities}
\label{ssec:int_mag}
\begin{figure*}
\centering    
\includegraphics[width=\textwidth]{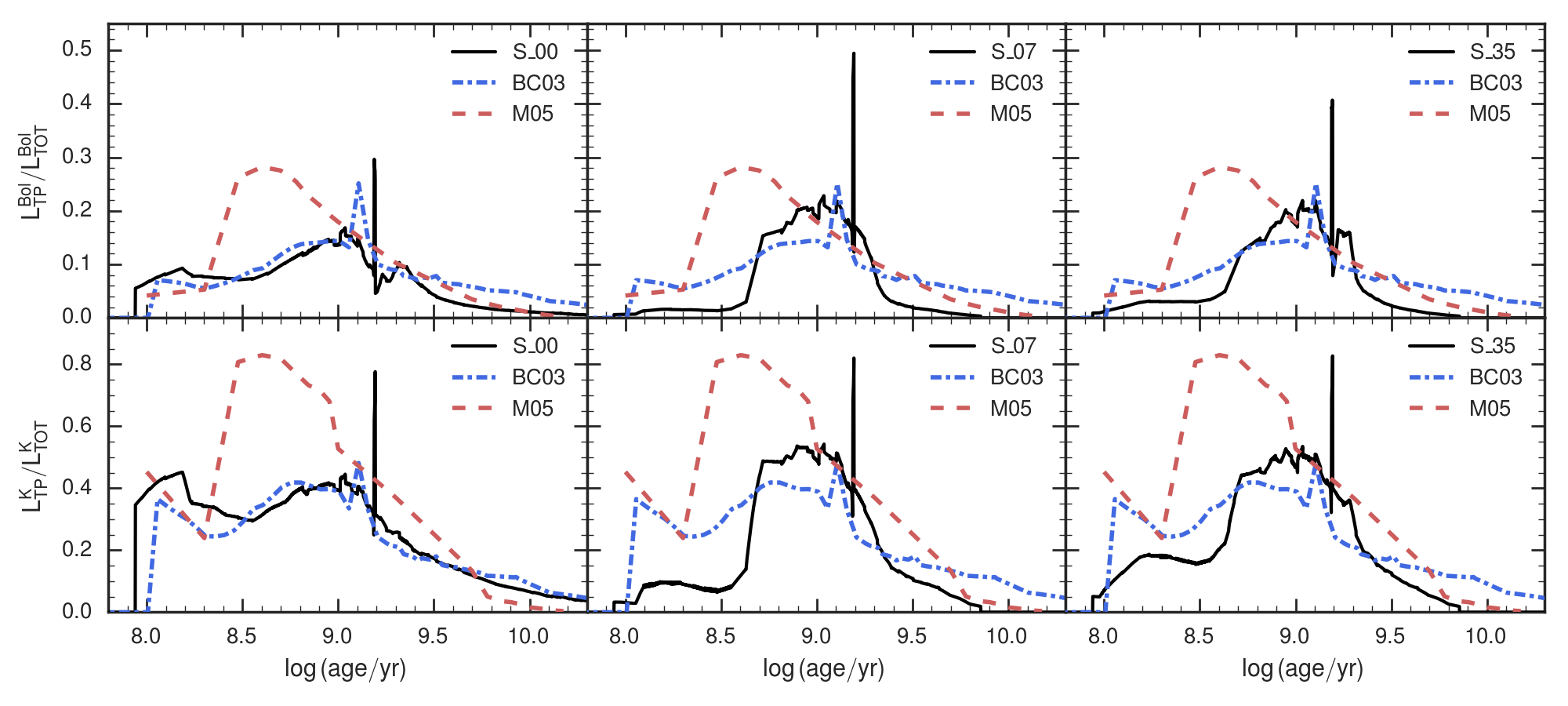}
\caption{Contribution of TP-AGB stars to the total bolometric luminosity (upper panels) and the K-band luminosity (lower panels), as a function of age ($\Zini = 0.004$), as predicted by the initial set S\_00 (left panels), and the two best-fitting sets S\_07 (middle panels) and S\_35 (right panels). The results by \citet[][M05]{Maraston_etal_05} and \citet[][BC03]{BC03} are shown as red dashed lines and blue dashed-dotted lines respectively. }
\label{fig:int_mags}
\end{figure*}

One of the most important motivations for calibrating TP-AGB evolutionary tracks, resides in their contribution to the integrated light, or spectral energy distribution (SED), of unresolved galaxies. Indeed, integrated spectra are often the only source of information about the total stellar masses, mean ages and metallicities of galaxies. Since \citet{Maraston_etal_05}, TP-AGB stars have been recognized as a crucial contributor to the light of galaxies, especially at (rest-frame) infrared wavelengths, and for galaxies containing $\sim1$-Gyr populations, at which this contribution is expected to peak. A few subsequent works attributed a much smaller importance to TP-AGB stars \citep[e.g.][]{conroy10,kriek10,Zibetti_etal_13}, lowering down their  contribution to values similar to those presented in the classical population synthesis models by \citet{BC03}. Entering into the details and consequences of this ongoing debate \citep[e.g.][]{maraston13} is far beyond the scope of this paper; anyhow let us check what amount of TP-AGB contribution is favoured by our calibration.

Fig.~\ref{fig:int_mags} displays the expected contribution of TP-AGB stars to the total integrated bolometric luminosity and the $K$-band luminosity as a function of age for a stellar population of \Zini = 0.004 for the initial set S\_00 and the two best-fitting models S\_07 and S\_35. We also show the contribution of TP-AGB stars as derived by \citet[][private communication]{BC03} and \citet[][private communication]{Maraston_etal_05} for the same metallicity\footnote{\citet{Maraston_etal_05} does not include the results for metallicity \Zini = 0.004. We used a linear interpolation between the models with \Zini = 0.001 and \Zini = 0.02.}.
  
We compute the integrated luminosities assuming an age interval of $\log(\mathrm{age/yr}) = 7.5 - 10.5$, with a step of 0.001. The spike visible at $\log(\mathrm{age/yr}) \sim 9.2 $ is caused by the TP-AGB boosting effect, already mentioned in Section~\ref{ssec:tri_sim} and fully described by \citet{girardi13}.

The results for the two best-fitting models are very similar. They differ mainly in the contribution of the youngest TP-AGB stars ($\log(\mathrm{age/yr}) \lesssim 8.5$), which is slightly lower in S\_07.
The lower TP-AGB contribution for $\log(\mathrm{age/yr}) \lesssim 8.8$ and $\log(\mathrm{age/yr}) \gtrsim 10$ predicted by S\_07 and S\_35 with respect to S\_00 is a consequence of the reduced lifetimes of the more massive stars ($\Mini \gtrsim 3~\Msun$) and the low-mass O-rich stars respectively.

In general, our calibration favours a mild TP-AGB contribution that does not exceed $\simeq$ 50 per cent in the $K$-band luminosity ($\simeq$ 20 per cent in bolometric luminosity)
for ages between $\sim 0.5$ and $\sim 2$~Gyr. These results are closer to the BC03 models, whereas M05 models show a larger contribution of the TP-AGB, up to 80 per cent in the $K$-band luminosity ($\simeq$ 30 per cent in bolometric luminosity). We note that a sizable reduction of the TP-AGB contribution in M05 models is also suggested by \citet{noel_et_al_2013}, based on a re-analysis of the integrated colours of Magellanic Clouds' clusters.

\subsection{Characterization of AGB star population}
\label{ssec:characterization}

\begin{figure*}
\centering    
\includegraphics[width=\textwidth]{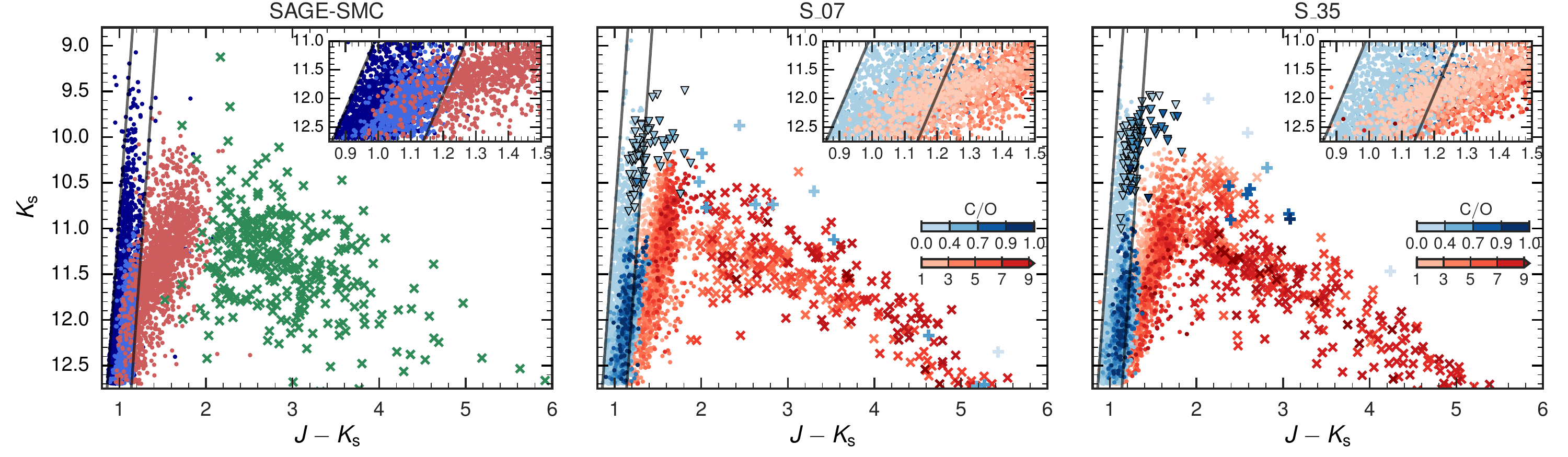}
\caption{Left panel: observed \cmd{\ks}{J}{\ks} CMD with stars colour-coded according to the B11 and SR16 classification (O-rich in blue, a-AGB in light blue, C-rich in red and X-AGB stars are shown with green crosses). Middle and right panels: simulated CMDs from the best-fitting set S\_07 and S\_35 with stars colour-coded according to the predicted \co\ ratio (O-rich stars are overplotted to C-rich stars). HBB stars are shown with triangles and with plus symbols when they fall in the X-AGB classification criteria. The insets show the CMD region where the O-rich and C-rich stars cannot be clearly separated using the classical photometric criteria shown as solid lines \citep{cioni06a,boyer11,boyer15} (C-rich stars are overplotted to O-rich stars).}
\label{fig:obs_S35_co}
\end{figure*}

Given our detailed population synthesis simulations we can provide a full characterization of the AGB population in the SMC in terms of stellar parameters. A comparison between the best-fitting models S\_07 and S\_35 and the observations in the \cmd{\ks}{J}{\ks} CMD is shown in Fig.~\ref{fig:obs_S35_co}. Both sets of models give essentially the same results in terms of the predicted stellar photometry. Most of the stars classified as X-AGB are C-rich (in agreement with the results from the SED fitting by SR16), with a small contamination of O-rich stars undergoing HBB. The stars populating the upper part of the CMD, with \ks\ magnitudes brighter than 10 mag, are exclusively O-rich and most of them are experiencing HBB. The inset of Fig.~\ref{fig:obs_S35_co} clearly shows that C-rich and O-rich stars cannot be accurately separated by using classical \jks\ criteria, in agreement with \citet{boyer15}.

\begin{figure*}
    \centering
    \includegraphics[width=\textwidth]{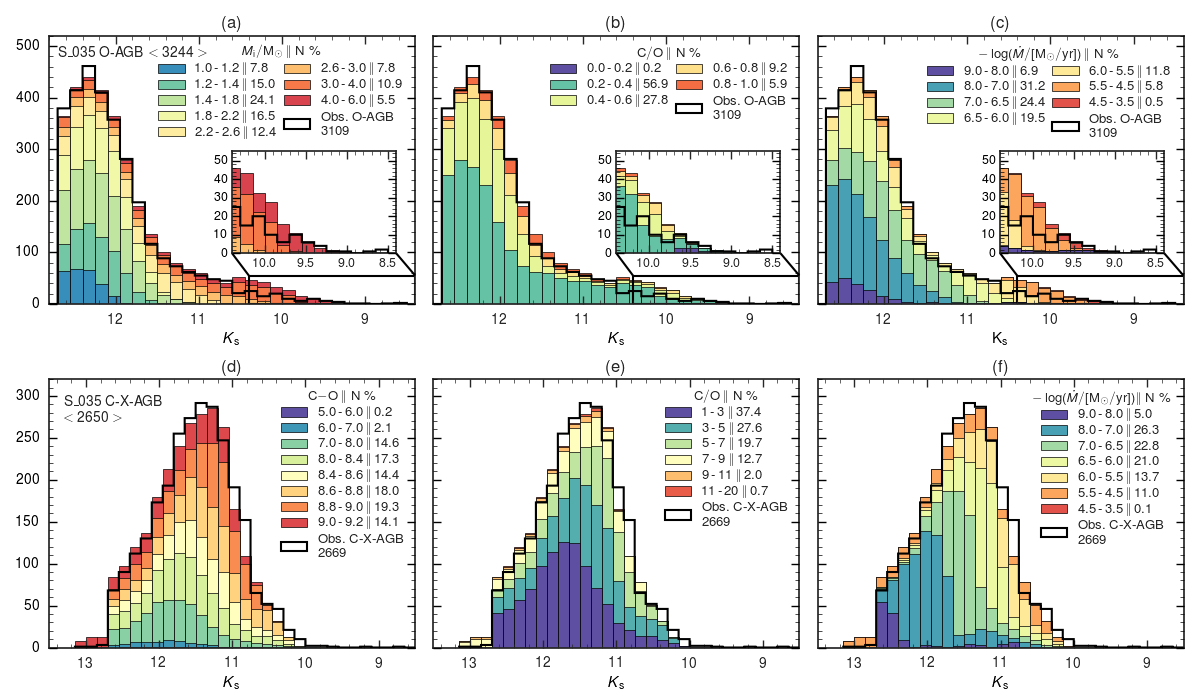}
    \caption{\ks-band LFs from the best-fitting set S\_35 decomposed in bins of selected stellar parameters (\Mini, \cminuso, \co, and \mdot) as indicated in each panel. The O-rich LFs are shown in the three upper panels, whereas the lower panels shows the LFs of the C- plus X-AGB.
    The observed LFs are shown as solid black lines. The legend of each panel shows the selected bins and the corresponding percentage of synthetic stars. The synthetic LFs are constructed as the average of the 10 \trilegal\ realisations.}
    \label{fig:SMC_stack}
\end{figure*}

A more quantitative analysis can be performed with the help of the \ks-band LFs decomposed in bins of selected stellar parameters. In Fig.~\ref{fig:SMC_stack}, we show some of the most relevant ones for the O-rich and the C- and X-AGB  populations (i.e. \co, \cminuso, \mdot\ and \Mini) as predicted by the best-fitting set S\_35. 
The bulk of O-rich stars contains low-mass stars with $\Mini \lesssim 2$~\Msun\ and \ks\ magnitudes fainter than $\approx 11$ mag. Most massive TP-AGB stars mainly populate the bright-end of the LF (see panel (a) of Fig.~\ref{fig:SMC_stack}).
As shown in panel (b) of Fig.~\ref{fig:SMC_stack}, the predicted values of the C/O ratios are between 0.2 and 0.6 for the $\approx 80$ per cent of the stars. The contribution of stars that have a lower surface abundance of carbon as a consequence of the HBB is visible at $\ks \approx 9.5$ mag. As for the mass-loss rates attained by O-rich stars, the models predict a rate around  $10^{-7}$~\MsunYr\ for a half of them, and the higher mass-loss rates, i.e. $\mdot \gtrsim 10^{-5}$~\MsunYr, are predicted for the most massive and luminous stars (see panel (c) of Fig.~\ref{fig:SMC_stack}). 
A fundamental parameter for the C-rich stars is the \cminuso\ as it is the main parameter that determines the activation of the CDYN \mdotdust, the minimum \cminuso\ value being $\approx 8.2$. 
The bulk of C-stars has $\cminuso \gtrsim 8$ and values larger than 9 are attained by the X-AGB stars, as can be appreciated in panel (d) of Fig.~\ref{fig:SMC_stack}. 
The predicted values of the C/O ratio are also shown in panel (e) of  Fig.~\ref{fig:SMC_stack}  and they essentially reflect the distribution of \cminuso\ values. Finally, the predicted mass-loss rates for the C-rich stars are similar to those of the O-rich, with the X-AGB stars reaching the higher mass-loss rates (see panel (f) of Fig.~\ref{fig:SMC_stack}). 

\begin{figure}
\centering
\includegraphics[width=\columnwidth]{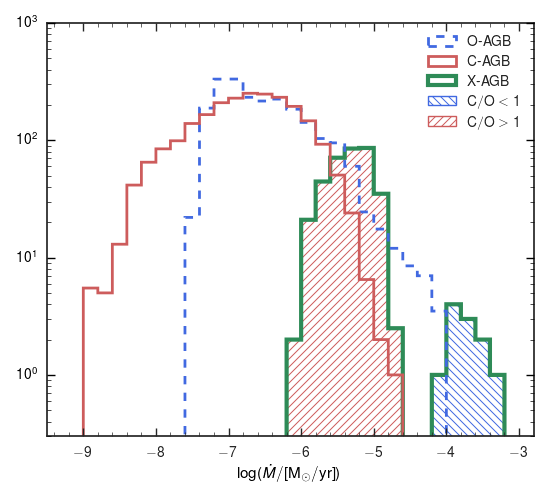}
\caption{Distributions of the predicted mass-loss rates for the C-, X- and O-AGB synthetic populations from the best-fitting model S\_35. Only stars in the TP-AGB phase are included. The X-AGB stars with $\co>1$ and those with $\co<1$ are shown as red and blue hatched regions, respectively, as indicated in the legend. }
\label{fig:mloss_distrib}
\end{figure}

The distributions of the predicted mass-loss rates for the C-, X- and O-AGB synthetic population are shown Fig.\ref{fig:mloss_distrib}.    
The predicted mass-loss rates for both O-rich and C-rich stars are in agreement with the mass-loss rates estimates derived from the spectral energy distribution (SED) fitting performed by \citet{nanni18} for the same sample of C- and X-AGB stars used in this work. 
In particular, the separation between C- and X-AGB stars at $\jks\ \approx 2$ occurs when the mass-loss rate reaches $\approx  10^{-6}$~\MsunYr\ (see Fig.~\ref{fig:obs_S35_co}), in agreement with the value found by \citet{nanni18}.
The predicted ranges of mass-loss rates are also in agreement with the results of three additional SED fitting studies of AGB stars in the SMC by SR16, \citet{groenewegen_sloan_18} and \citet{goldman_etal_18}.

\begin{figure}
\centering
\includegraphics[width=\columnwidth]{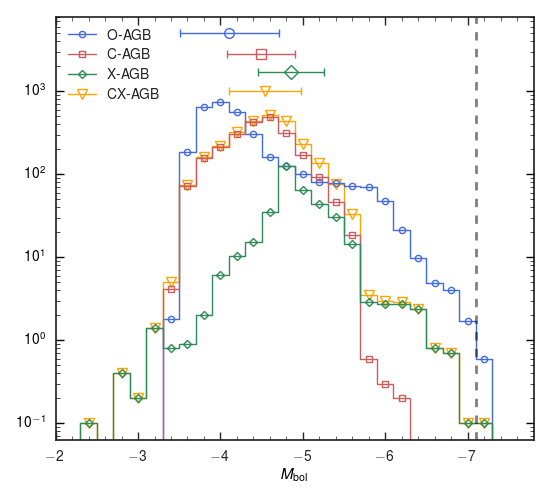}
\caption{Bolometric magnitude distributions of the synthetic populations of O-, C and X-AGB stars as predicted by the best-fitting set S\_35. The distribution for the combined C- and X-AGB populations is also shown.
The median values and the $1\sigma$ dispersion of the bolometric magnitude distributions are  $-4.11 \pm 0.60$, $-4.49 \pm 0.43$, $-4.86 \pm 0.40$ and $-4.54 \pm 0.44$ mag for the O-, C, X-AGB and C- and X-AGB combined respectively, as also shown in the upper part of the figure. The vertical dashed line is the ``classical'' AGB limit at \Mbol = -7.1 mag. The distributions are calculated as the median of the 10 \trilegal\ simulations.  }
\label{fig:mbol}
\end{figure}

It is also interesting to compare the distributions of the bolometric magnitudes (\Mbol)  as predicted by the best-fitting model S\_35 with the results of SR16, B11 and \citet{nanni18}. SR16  estimated the luminosities of the observed AGB stars in the SMC using the results of their best-fitting SED models, whereas B11 performed a trapezoidal integration from the observed optical $U$-flux through the mid-infrared 24 $\mu$m-flux. \citet{nanni18} performed a similar study on the sample of C-rich and X-AGB stars identified by SR16.
The predicted \Mbol-LFs as derived from the average LFs of the 10 \trilegal\ simulations with the TP-AGB set S\_35 are shown in Fig.~\ref{fig:mbol}.
The predicted peaks of the \Mbol\ distributions, i.e. the median values,  of the O-rich, C-rich and X-AGB LFs are $-4.11$, $-4.49$ and $-4.86$ mag, respectively. The combined C- and X-AGB LF peaks at $-4.54$ mag. 
The values reported by B11 are $-4.59$ mag for the C-AGB and $-4.63$ mag for the C- and X-AGB samples combined, and the values estimated by SR16 are $-4.2$ mag for the O-rich and $-4.5$ mag for the C-rich stars (including the stars classified as X-AGB). The bolometric luminosity distribution of the C-rich and X-AGB stars estimated by \citet{nanni18} are very similar and are well reproduced by our best-fitting model. The predicted bolometric luminosities are in excellent agreement with both the results of B11 and SR16, given that small discrepancies can be due to the fact that both authors adopted a fixed distance modulus across the SMC.

\subsection{2MASS and {\em Spitzer} luminosity functions}

\begin{figure*}
\centering
\includegraphics[width=\textwidth]{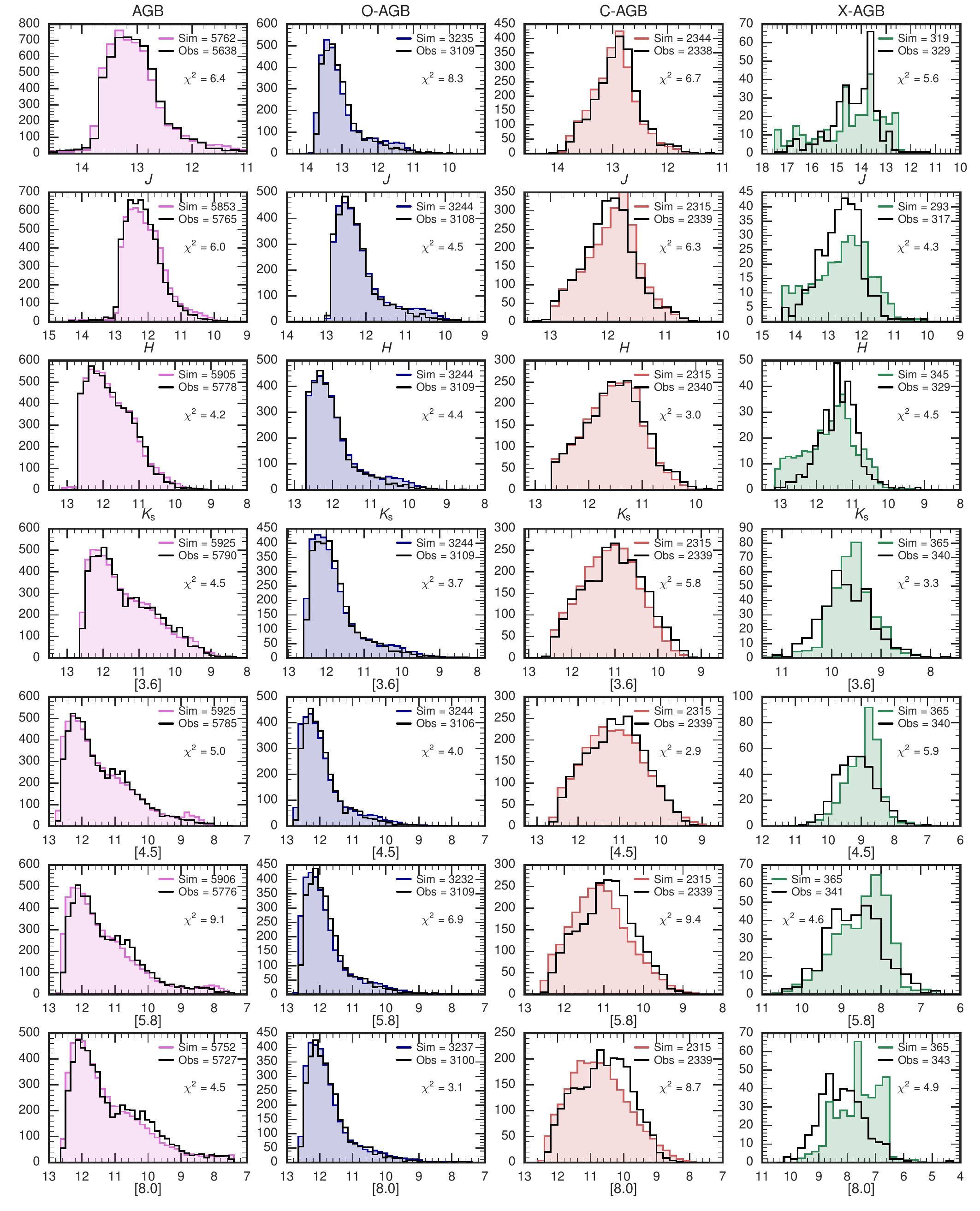}
\caption{Comparison between the synthetic LFs obtained from the best-fitting model S\_35 and the observed LFs in the 2MASS and Spitzer filters, going from shorter (top panels) to longer wavelengths (bottom panels). }
\label{fig:sage_lfs_S35}
\end{figure*}

Figure~\ref{fig:sage_lfs_S35} shows the comparison between our best-fitting model S\_35 and the LFs in the 2MASS and \textit{Spitzer} bands.
We find a general satisfactory  agreement between our best-fitting model and the  observed LFs, which supports the validity of our treatment of the circumstellar dust based on 
an improved version of \citet{marigo08}.
A few discrepancies exist, in particular affecting the LFs of X-AGB class in some photometric bands. This is the case of the \textit{Spitzer} filters [4.5], [5.8] and [8.0]; the predicted X-AGB LFs are shifted to brighter magnitudes with respect to the observed ones. At the same time, the predicted C-AGB LFs in the \textit{Spitzer} filters [5.8] and [8.0] appear slightly shifted towards magnitudes fainter than observed.
Overall, these differences are of small entity, and more importantly, we expect they 
do not impact on the results of our TP-AGB calibration, since the X-AGB sample includes  less than 6 per cent of the total number of AGB stars.

\section{Summary}
\label{sec:sum}

In this study we present and discuss a detailed calibration of the TP-AGB phase, focusing on resolved stellar populations in the Small Magellanic Cloud.
To this aim, we couple high-quality observational data (from the optical to the mid-infrared) to detailed models for AGB stars.
Several ingredients are put in place, namely:  the spatially resolved SFH and AMR of the SMC derived from the deep near-infrared photometry of the VMC survey \citep{rubele18}; the complete census of TP-AGB stars in the SMC identified in the 2MASS and \textit{Spitzer} pass-bands, together with their classification (O-rich, C-rich, extreme-AGB stars) contained in the catalogues by \citet{boyer11} and \citet{SR16}; large grids of stellar evolutionary tracks computed with the \parsec\ code \citep{bressan12}, corrected by the early-AGB mass loss, and coupled with detailed TP-AGB evolutionary tracks computed with the \colibri\ code \citep{marigo13}; state-of-the-art stellar spectral libraries for O-rich \citep{aringer16} and C-rich \citep{aringer09} stars; a treatment to account for the effect of circumstellar dust on the photospheric emission based on radiative transfer calculations \citep[an improved version of][]{marigo08}; new pulsation models for long-period variables \citep{Trabucchi_etal18}; extended grids of stellar isochrones with rich information about the properties of AGB stars (modulation of luminosity and effective  temperature during the thermal pulse cycles, surface chemical abundances) following the implementation described in \citet[][]{marigo17}; a well-tested stellar population synthesis code \citep[][\trilegal]{girardi05} that includes a proper estimation of photometric errors and completeness.

We produced a large grid  of TP-AGB models  with the aim of identifying suitable combinations of mass loss and third dredge-up laws that best fit a set of observables at the same time, namely: the \ks-band luminosity functions of the observed O-rich, C-rich and extreme-AGB stars.
Among the many computed sets, two are found to recover the observational constraints with comparable 
performance. Both sets share the same prescriptions for the pre-dust mass loss \citep{CranmerSaar_11}, and the dust-driven mass loss of C-rich stars \citep{Eriksson_etal14, mattsson10}.
The main differences between them refer to the adopted efficiency of the dust-driven winds during the O-rich stages, and the description of the 3DU.

The first best-fitting set describes the efficiency $\lambda$ of the 3DU following the  formalism introduced by \citet[][]{karakas02} on the basis of their full TP-AGB models. Given the high efficiency ($\lambda \simeq 1$) predicted by this formalism for $\Mini \ga 3 $~\Msun, the only way to avoid a sizeable excess of C-stars at magnitudes $\ks < 10$ mag is to invoke a strong mass loss during the O-rich stages that precede the formation of C-stars  \citep[][with $\etadust = 0.06$]{bloecker95}. In this way, for stars with $3~\Msun \la \Mini \la 4~\Msun$ (at $\Zini=0.004$) the complete ejection of the envelope takes place before the 3DU is able to increase the C/O ratio above unity. For $\Mini \ga 4~\Msun$  the lack of C-stars is also due to the concomitant occurrence of HBB which consumes carbon in favour of nitrogen.

The second  best-fitting set adopts a new formalism for the 3DU. Starting from the results of full AGB models in the literature \citep{VenturaDantona_09, Cristallo_etal15}, we build a parametrization  in which $\Lambdamax$ (the maximum value attained during the TP-AGB phase) first increases with the mass of the star, reaches a maximum value and then decreases in the most massive AGB stars. 
Our calibration indicates that at $\Zini=0.004$ the efficiency of the 3DU increases up to about $\Lambdamax \simeq 0.6$ in stars with $\Mini \simeq 2.0-2.5$~\Msun, and then decreases at larger masses, down to $\Lambdamax \la 0.2$ for $\Mini > 5$~\Msun. We note that such decline of $\Lambdamax$, for $\Mini > 4$~\Msun in particular, is at odds with the 3DU properties of the TP-AGB models by  \citet[][]{karakas02}, while appears in line with the trends predicted by the TP-AGB models of \citep{VenturaDantona_09} and \citet{Cristallo_etal15}.
Thanks to the reduction of $\Lambdamax$ at larger masses, the second best-fitting set avoids naturally the excess of bright carbon stars, so that the O-rich stages can be suitably described with a lower efficiency parameter for the dust-driven mass loss \citep[][with $\etadust = 0.02-0.03$]{bloecker95}. 

The two sets produce similar star counts for all classes of AGB stars in the SMC,  recover their $\ks$-band luminosity functions  quite well, and yield comparable contributions to the integrated light of simple stellar populations as a function of the age. The comparison with the  \textit{Spitzer} observations in the mid-infrared bands is also satisfactory, indicating that the treatment of the bolometric corrections for the circumstellar dust around mass-losing AGB stars is reasonable.

The similarly good performance of both sets suggests that the final results are affected by a subtle degeneracy. However, this latter can be resolved by considering complementary predictions for AGB  stars, namely the IFMR and the AGB chemical ejecta.
As to the IFMR,
we find that, compared to the S\_07 set, the second best-fitting set S\_35, characterized by a reduced efficiency of the 3DU for $\Mini > 3$~\Msun, at solar metallicity predicts higher white dwarf masses, getting in better agreement with the semi-empirical relations newly derived by \citet{gaia_ifmr}  and \citet[][]{cummings_18}.  Based on this result, we may conclude that the set S\_035 is to be preferred.

The second aspect that is relevant to further probe the models is related to the chemical yields of AGB stars. In particular, a careful examination of the surface elemental abundances (e.g., lithium, CNO elements, sodium, magnesium, aluminum, s-process elements) in high-luminosity AGB stars that experience HBB may help to discriminate between the two sets. We postpone a detailed investigation to a follow-up work.

\section{Conclusions}
\label{sec:concl}

The main conclusions we draw from the results of the present work can be outlined as follows: 
\begin{itemize} 
\item We confirm the importance of the pre-dust mass loss for the low-mass (with initial masses around 1~\Msun), low-metallicity O-rich stars, as already found by earlier works \citep{girardi10, rosenfield14, rosenfield16}.

\item The inclusion of a mass-loss prescription based on detailed dynamical models for C-stars \citep{Eriksson_etal14, mattsson10} is of key importance to correctly describe the population of this class of TP-AGB stars, given its dependence on stellar parameters, in particular the carbon excess. 

\item The minimum  and maximum initial masses for the formation of C-stars are around 1.4~\Msun\ and 3~\Msun\,
respectively, at SMC-like metallicity (i.e. $\Zini=0.004$) that is representative 
for the bulk of C-stars.  
The lack of C-stars for $\ks < 10$ mag clearly indicates that the third dredge-up in TP-AGB stars with $\Mini \ga 3~\Msun$ should not take place over many thermal pulses, nor with extreme efficiency.
The minimum mass threshold is found to increase with the initial metallicity, in agreement with previous studies \citep[e.g.][]{marigo07}.
The maximum mass threshold also shows a positive correlation with metallicity, though less pronounced.
In general, these trends would lead the C/M ratio to decrease in host systems of increasing metallicity (depending also on their SFH), a well-known fact in the literature \citep[e.g.][]{boyer13}.

\item The number counts of the brightest O-rich stars experiencing HBB can be reproduced by invoking a mass loss with a significant luminosity dependence \citep[][with $\etadust=0.02-0.03$ for a shallow 3DU, or with $\etadust=0.06-0.07$ for an extremely deep 3DU]{bloecker95}.
We find that a good reproduction of the O-rich LF is also attained with the empirical relation between mass-loss rate and pulsation period proposed by \citet{VassiliadisWood_93}.

\item A direct outcome of our calibration, based on resolved stellar populations, is the full characterization of the AGB population in the SMC in terms of stellar parameters (i.e. initial mass, mass-loss rate, C/O, carbon excess, luminosity).
The distribution of mass-loss rates for C-rich and extreme-AGB stars agree well with the one derived by \citet{nanni18} through the technique of SED fitting  applied to the same observed catalogue adopted in this work. We also find a good agreement with the mass-loss rate estimates derived by \citet{groenewegen_sloan_18} and \citet{goldman_etal_18}.
The main features (the position of the peak and the width) of the predicted distributions in bolometric  magnitude for the O-rich, C-rich and extreme-AGB star classes are in line  with studies based on the SED-fitting methodology \citep{SR16}, as well as with the work of \citet{boyer11}.

\item Considering the characteristics of the method and the data employed in the calibration, it is fair to state that some conclusions might be more robust than others. The most reliable constraints apply to the lifetimes and the bolometric luminosities of the different TP-AGB subtypes (especially Figs.~\ref{fig:mbol_tr_lum} and \ref{fig:lifetimes}), at SMC metallicities. These properties are directly related to the observational LFs being fitted, with some uncertainty coming mainly from the limited resolution in initial masses inherent to our method. Lifetimes and bolometric luminosities are also linked to the contribution of TP-AGB stars to the integrated SED of galaxies. Our results, obtained at SMC metallicities, appear to support a TP-AGB contribution intermediate between those favoured by the popular models from \citet{BC03} and \citet{Maraston_etal_05} -- although pending towards the ``light contribution'' side given by \citet{BC03} models. In this respect additional indications  may be derived in the near future, through the analyses of other, more metal-rich galaxies (namely the LMC and M31) using the same method.

\end{itemize}

While the present calibration covers the metallicity range defined by the AMR of the SMC ($0.0005 \lesssim Z\lesssim 0.008$), we extend the calculation of the TP-AGB tracks to cover the interval $0.0005<Z<0.03$ ($-1.5 \lesssim \feh \lesssim +0.3$).
Extended \texttt{PARSEC+COLIBRI} isochrones derived from these tracks (of both sets S\_07 and S\_35) can be retrieved from our web interfaces \url{http://starkey.astro.unipd.it/cmd} and \url{http://stev.oapd.inaf.it/cmd}, for over 50 different photometric systems including major photometric surveys and instruments (e.g.\ 2MASS, SDSS, \textit{Spitzer}, AKARI, HST/ACS, HST/WFC3, WISE, VISTA, \textit{Gaia}, TESS), and future datasets (e.g.\ JWST, LSST, Euclid). The isochrones also
include pulsation properties of long-period variables according
to the prescriptions of \citet{Trabucchi_etal18}.
A general description of the isochrone characteristics and data format is provided in \citet{marigo17}.

\section*{Acknowledgements}
This work is supported by the ERC Consolidator Grant funding scheme ({\em project STARKEY}, G.A. n. 615604). 
We thank the entire VMC team for producing the space-resolved SFH used in this work;  J.~Cummings and J.~Kalirai for kindly providing us with their IFMR data.
MRC acknowledges funding from the European Research Council (ERC) under the European Union's Horizon 2020 research and innovation programme (grant agreement No. 682115).
Many thanks go to C.~Maraston, S.~Charlot, G.~Bruzual for providing us with their stellar population synthesis models.
This publication makes use of data products from the Two Micron All Sky Survey, which is a joint project of the University of Massachusetts and the Infrared Processing and Analysis Center/California Institute of Technology, funded by the National Aeronautics and Space Administration and the National Science Foundation.
This research made use of Astropy, a community-developed core Python package for Astronomy \citep{astropy13, astropy18} and matplotlib, a Python library for publication quality graphics \citep{matplotlib}.

\bibliographystyle{mnras}



\appendix

\input{appendixA}

\input{appendixB}

\bsp	
\label{lastpage}
\end{document}

%% file: macros/newcommands.tex
\newcommand {\reac}[6] {$\rm\,{}^{#2}\kern-0.8pt{#1}\,({#3}\,,{#4})  \,{}^{#6}\kern-0.8pt{#5}\,$}
\newcommand{\sqdeg}{\mbox{deg$^{2}$}}

\newcommand{\Mini}{\mbox{$M_{\rm i}$}}
\newcommand{\Zini}{\mbox{$Z_{\rm i}$}}

\newcommand{\jks}{\mbox{$J\!-\!K_{\rm s}$}}
\newcommand{\yks}{\mbox{$Y\!-\!K_{\rm s}$}}

\newcommand{\ks}{\mbox{$K_{\rm s}$}}

\newcommand{\col}[2]{\mbox{$#1 \!-\! #2$}}
\newcommand{\cmd}[3]{\mbox{$#1$ vs.\ $#2\!-\!#3$}}

\newcommand{\Mbol}{\mbox{$M_{\rm bol}$}}

\newcommand{\av}{\mbox{$A_V$}}

\newcommand{\feh}{\mbox{\rm [{\rm Fe}/{\rm H}]}}
\newcommand{\mh}{\mbox{\rm [{\rm M}/{\rm H}]}}
\newcommand{\Msun}{\mbox{$\mathrm{M}_{\odot}$}}

\newcommand{\Teff}{\mbox{$T_{\rm eff}$}}

\newcommand{\co}{\mbox{${\rm C/O}$}}
\newcommand{\cminuso}{\mbox{$\mathrm{C}-\mathrm{O}$}}

\newcommand{\mdot}{\mbox{$\dot {M} $}}

\newcommand{\chisqlf}{\mbox{$\chi^2_{\rm LF}$}}

\newcommand{\Lambdamax}{\mbox{$\lambda_\mathrm{max}$}}
\newcommand{\Tbdred}{\mbox{$T_{\rm b}^{\rm dred}$}}
\newcommand{\Mcmin}{\mbox{$M_{\rm c}^{\rm min}$}}

\newcommand{\ETA}{\mbox{$\eta$}}
\newcommand{\etapre}{\mbox{$\eta_{\rm pre-dust}$}}
\newcommand{\etadust}{\mbox{$\eta_{\rm dust}$}}
\newcommand{\mdotpre}{\mbox{$\dot {M}_{\rm pre-dust}$}}
\newcommand{\mdotdust}{\mbox{$\dot {M}_{\rm dust}$}}

\newcommand{\trilegal}{\texttt{TRILEGAL}}
\newcommand{\parsec}{\texttt{PARSEC}}
\newcommand{\colibri}{\texttt{COLIBRI}}
\newcommand{\MsunYr}{\mbox{$\mathrm{M}_{\odot }\,\mathrm{yr}^{-1}$}} 

\newcommand{\beq}{\begin{equation}}
\newcommand{\eeq}{\end{equation}}
\newcommand{\beqa}{\begin{eqnarray}}
\newcommand{\eeqa}{\end{eqnarray}}

%% file: tpagb_grid.tex
\begin{tabular}{c|cc ccccc| cccc } 
\hline
\hline 
 & \multicolumn{7}{c|}{\textbf {Mass Loss}} & 
   \multicolumn{4}{c}{\textbf{Third Dredge-Up}} \\
 
 & \multicolumn{2}{c}{ \textbf{Pre-Dust}} & 
   \multicolumn{4}{c}{ \textbf{Dust - driven}}  & &
   
   \textbf{Activation} &  
   \multicolumn{3}{c}{\textbf{Efficiency}} \\             

 &  &  & \multicolumn{2}{c}{ \textbf{M-stars}}  &
 \multicolumn{2}{c}{ \textbf{C-stars}}        &  &  &  & &  \\ 
   
\textbf{SET}  &  \textbf{Id} & \textbf{\ETA}  & \textbf{Id}  & \textbf{\ETA} & \textbf{Id}  & \textbf{\ETA} & $\dot{M}^{(1)}$ &  $\log(\Tbdred/{[\rm K]})^{(2)}$   
&  \multicolumn{3}{c}{ \textbf{$\lambda$}} \\ 

\hline
S\_00 & SC05 & -- & BE88 & -- & BE88 & --    & (a) & 6.40 &
\multicolumn{3}{c}{ K02} \\ 
S\_01 & CS11 & 2 & BL95 & 0.05 & BL95 & 0.05 & (a) & 6.40 & 
\multicolumn{3}{c}{ K02} \\ 
S\_02 & CS11 & 2 & BL95 & 0.02 & CDYN & 1    & (a) & 6.40 & 
\multicolumn{3}{c}{ K02} \\ 
S\_03 & CS11 & 2 & BL95 & 0.03 & CDYN & 1    & (a) & 6.40 &  
\multicolumn{3}{c}{ K02} \\ 
S\_04 & CS11 & 2 & BL95 & 0.05 & CDYN & 1    & (a) & 6.40 &
\multicolumn{3}{c}{ K02} \\ 
S\_05 & CS11 & 2 & BL95 & 0.06 & CDYN & 1    & (a) & 6.40 &
\multicolumn{3}{c}{ K02} \\ 
S\_06 & CS11 & 3 & BL95 & 0.06 & CDYN & 1    & (a) & 6.40 &
\multicolumn{3}{c}{ K02} \\ 
S\_07 & CS11 & 3 & BL95 & 0.06 & CDYN & 1    & (a) & f$_{1}$(\Zini) & 
\multicolumn{3}{c}{ K02} \\ 
S\_19 & CS11 & 3 & VW93 & --   & CDYN & 1    & (a) & f$_{1}$(\Zini) &
\multicolumn{3}{c}{ K02} \\
S\_12 & CS11 & 3 & BL95 & 0.01 & CDYN & 1    & (b) & f$_{1}$(\Zini) &
\multicolumn{3}{c}{ K02} \\ 
S\_08 & CS11 & 3 & BL95 & 0.06 & CDYN & 1    & (a) & f$_{1}$(\Zini) & \multicolumn{3}{c}{ \Lambdamax = 0.5} \\ 
S\_09 & CS11 & 3 & BL95 & 0.02 & BL95 & 0.02 & (a) & f$_{1}$(\Zini) & \multicolumn{3}{c}{ \Lambdamax = 0.5} \\ 
S\_10 & CS11 & 3 & BL95 & 0.01 & BL95 & 0.01 & (a) & f$_{1}$(\Zini) & \multicolumn{3}{c}{ \Lambdamax = 0.5} \\ 
S\_11 & CS11 & 3 & BL95 & 0.01 & CDYN & 1    & (b) & f$_{1}$(\Zini) & \multicolumn{3}{c}{ \Lambdamax = 0.5} \\ 
S\_31 & CS11 & 3 & BL95 & 0.01 & CDYN & 1    & (b) & f$_{1}$(\Zini) & \multicolumn{3}{c}{\Lambdamax = 0.4} \\
      &  &   &  & &  &  &  & &
      $\lambda_{\rm max}^{\ast}$ &
      $\mathrm{\widetilde{M}_c}[\Msun]$      &   
      $\mathrm{M_{c,\lambda=0}[\Msun]}$ \\
\cline{10-12}
S\_13 & CS11 & 3 & BL95 & 0.01 & CDYN & 1 & (b) & f$_{1}$(\Zini) & 0.5 & 0.65 & 0.95 \\
S\_14 & CS11 & 3 & BL95 & 0.01 & CDYN & 1 & (b) & f$_{1}$(\Zini) & 0.5 & 0.65 & 0.85 \\ 
S\_15 & CS11 & 3 & BL95 & 0.01 & CDYN & 1 & (b) & f$_{1}$(\Zini) & 0.6 & 0.60 & 0.85 \\
S\_16 & CS11 & 3 & BL95 & 0.01 & CDYN & 1 & (b) & f$_{2}$(\Zini) & 0.6 & 0.60 & 0.85 \\
S\_17 & CS11 & 3 & BL95 & 0.01 & CDYN & 1 & (c) & f$_{2}$(\Zini) & 0.6 & 0.60 & 0.85 \\
S\_18 & CS11 & 3 & BL95 & 0.02 & CDYN & 1 & (c) & f$_{2}$(\Zini) & 0.6 & 0.60 & 0.85 \\
S\_20 & CS11 & 3 & BL95 & 0.02 & CDYN & 1 & (a) & f$_{2}$(\Zini) & 0.6 & 0.60 & 0.85 \\ 
S\_22 & CS11 & 3 & BL95 & 0.02 & CDYN & 1 & (c) & f$_{2}$(\Zini) & 0.7 & 0.70 & 0.85 \\
S\_23 & CS11 & 3 & BL95 & 0.02 & CDYN & 1 & (c) & f$_{2}$(\Zini) & 0.8 & 0.60 & 0.85 \\
S\_24 & CS11 & 3 & BL95 & 0.02 & CDYN & 1 & (c) & f$_{2}$(\Zini) & 0.8 & 0.60 & 1.30 \\ 
S\_25 & CS11 & 3 & BL95 & 0.02 & CDYN & 1 & (c) & f$_{2}$(\Zini) & 0.8  & 0.60 & 1.00 \\ 
S\_26 & CS11 & 3 & BL95 & 0.02 & CDYN & 1 & (c) & f$_{2}$(\Zini) & 0.7  & 0.60 & 0.85 \\ 
S\_27 & CS11 & 3 & BL95 & 0.02 & CDYN & 1 & (c) & f$_{2}$(\Zini) & 0.7  & 0.60 & 1.00 \\ 
S\_28 & CS11 & 3 & BL95 & 0.02 & CDYN & 1 & (b) & f$_{2}$(\Zini) & 0.7  & 0.60 & 1.00 \\ 
S\_29 & CS11 & 3 & BL95 & 0.02 & CDYN & 1 & (c) & f$_{2}$(\Zini) & 0.7  & 0.625 & 1.00 \\
S\_30 & CS11 & 3 & BL95 & 0.02 & CDYN & 1 & (c) & f$_{2}$(\Zini) & 0.5  & 0.6 & 1.00 \\
S\_32 & CS11 & 3 & BL95 & 0.02 & CDYN & 1 & (c) & f$_{2}$(\Zini) & 0.5  & 0.5 & 1.00 \\
S\_34 & CS11 & 3 & BL95 & 0.02 & CDYN & 1 & (c) & f$_{2}$(\Zini) & 0.6  & 0.6 & 1.00 \\
S\_35 & CS11 & 3 & BL95 & 0.03 & CDYN & 1 & (c) & f$_{2}$(\Zini) & 0.7  & 0.60 & 1.00 \\ 

\hline
\multicolumn{12}{l}{ {\bf Notes:} } \\
\multicolumn{12}{l}{$^{(1)}$ Actual mass-loss rate:}\\
\multicolumn{12}{l}{ $\,\,\,\,\,$ (a) max(\mdotpre,\mdotdust)} \\
\multicolumn{12}{l}{ $\,\,\,\,\,$ (b) max(\mdotpre,\mdotdust), for $\co>1$ if \mdotdust is not active, we assume $\mdotdust = \mdotdust (\co <1)$ } \\ 
\multicolumn{12}{l}{ $\,\,\,\,\,$ (c) max(\mdotpre,\mdotdust) for $\co<1$; for $\co>1$ max(\mdotdust($\co<1$), \mdotdust($\co>1$) )} \\ 
\multicolumn{12}{l}{$^{(2)}$ 3DU activation temperature as a function of \Zini:}\\
\multicolumn{12}{l}{ $\,\,\,\,\,$ f$_{1}$(\Zini): $\log\Tbdred=\mathrm{max}[6.3, T_1+(T_2-T_1)(Z-Z_1)/(Z_2-Z_1)]$ with $T_1=6.3$, $T_1=6.60$, $Z_1=0.001$, $Z_2=0.02$.}\\
\multicolumn{12}{l}{ $\,\,\,\,\,$ f$_{2}$(\Zini): $\log\Tbdred=\mathrm{max}[6.2, T_1+(T_2-T_1)(Z-Z_1)/(Z_2-Z_1)]$ with $T_1=6.1$, $T_1=6.75$, $Z_1=0.001$, $Z_2=0.02$.}\\
\end{tabular}

%% file: SMC_dpmod_amc15sic_tab_results.tex
\begin{tabular}{r|ccc|ccc|ccc|ccc|c}
\hline
\hline
& \multicolumn{3}{c|}{AGB}
& \multicolumn{3}{c|}{O-AGB}
& \multicolumn{3}{c|}{C-AGB}
& \multicolumn{3}{c|}{X-AGB}
& C/M \\
& $N_\mathrm{TOT}$ & $\Delta N \% ^{1}$ & \chisqlf 
& $N_\mathrm{TOT}$ & $\Delta N \% ^{1}$ & \chisqlf 
& $N_\mathrm{TOT}$ & $\Delta N \% ^{1}$ & \chisqlf 
& $N_\mathrm{TOT}$ & $\Delta N \% ^{1}$ & \chisqlf 
&\\
\hline

OBS & 5778 & & &  3109 & & &  2340 & & &  329 & & &  0.86 \\
\hline
S\_00 & 6453 & 11.7 & 22.3 & 4484 & 44.2 & 20.5 & 1655 & -29.3 & 21.6 & 313 & -4.9 & 4.8 & 0.44 \\
S\_01 & 5385 & -6.8 & 16.0 & 3881 & 24.8 & 9.1 & 1335 & -42.9 & 71.9 & 168 & -48.9 & 6.5 & 0.39 \\
S\_02 & 7596 & 31.5 & 34.0 & 3756 & 20.8 & 7.3 & 3430 & 46.6 & 36.8 & 409 & 24.3 & 7.0 & 1.02 \\
S\_03 & 7395 & 28.0 & 27.2 & 3690 & 18.7 & 6.7 & 3321 & 41.9 & 30.4 & 383 & 16.4 & 5.0 & 1.0 \\
S\_04 & 6894 & 19.3 & 10.7 & 3682 & 18.4 & 5.8 & 2894 & 23.7 & 9.4 & 317 & -3.6 & 4.8 & 0.87 \\
S\_05 & 6713 & 16.2 & 8.2 & 3666 & 17.9 & 6.2 & 2755 & 17.7 & 7.2 & 290 & -11.9 & 5.2 & 0.83 \\
S\_06 & 5728 & -0.9 & 2.4 & 3155 & 1.5 & 2.6 & 2329 & -0.5 & 3.6 & 243 & -26.1 & 3.5 & 0.82 \\
S\_07 & 5805 & 0.5 & 2.3 & 3148 & 1.3 & 2.3 & 2380 & 1.7 & 4.0 & 276 & -16.1 & 3.5 & 0.84 \\
S\_08 & 5549 & -4.0 & 2.7 & 3193 & 2.7 & 2.6 & 2095 & -10.5 & 6.7 & 260 & -21.0 & 2.1 & 0.74 \\
S\_09 & 5521 & -4.4 & 4.9 & 3334 & 7.2 & 4.9 & 1990 & -15.0 & 6.7 & 197 & -40.1 & 5.1 & 0.66 \\
S\_10 & 5980 & 3.5 & 7.4 & 3387 & 8.9 & 6.1 & 2260 & -3.4 & 3.4 & 332 & 0.9 & 4.9 & 0.76 \\
S\_11 & 6072 & 5.1 & 12.6 & 3368 & 8.3 & 6.2 & 2366 & 1.1 & 9.1 & 337 & 2.4 & 4.5 & 0.8 \\
S\_12 & 6302 & 9.1 & 14.8 & 3317 & 6.7 & 4.3 & 2599 & 11.1 & 11.4 & 385 & 17.0 & 6.9 & 0.9 \\
S\_13 & 5882 & 1.8 & 12.7 & 3636 & 17.0 & 9.3 & 2014 & -13.9 & 14.9 & 231 & -29.8 & 5.6 & 0.62 \\
S\_14 & 5820 & 0.7 & 12.5 & 3774 & 21.4 & 13.6 & 1841 & -21.3 & 19.9 & 205 & -37.7 & 8.4 & 0.54 \\
S\_15 & 5843 & 1.1 & 10.9 & 3632 & 16.8 & 12.1 & 1952 & -16.6 & 9.6 & 258 & -21.6 & 3.4 & 0.61 \\
S\_16 & 5894 & 2.0 & 10.8 & 3491 & 12.3 & 10.5 & 2117 & -9.5 & 5.2 & 284 & -13.7 & 3.7 & 0.69 \\
S\_17 & 6602 & 14.3 & 18.2 & 3486 & 12.1 & 10.5 & 2792 & 19.3 & 11.3 & 323 & -1.8 & 6.2 & 0.89 \\
S\_18 & 6134 & 6.2 & 6.3 & 3407 & 9.6 & 6.2 & 2398 & 2.5 & 2.9 & 328 & -0.3 & 5.3 & 0.8 \\
S\_19 & 6542 & 13.2 & 29.7 & 3211 & 3.3 & 3.4 & 2948 & 26.0 & 32.2 & 382 & 16.1 & 8.0 & 1.04 \\
S\_20 & 5991 & 3.7 & 10.2 & 3436 & 10.5 & 6.5 & 2241 & -4.2 & 7.8 & 314 & -4.6 & 4.2 & 0.74 \\
S\_22 & 5882 & 1.8 & 7.5 & 4132 & 32.9 & 19.2 & 1508 & -35.6 & 59.6 & 242 & -26.4 & 4.0 & 0.42 \\
S\_23 & 6351 & 9.9 & 7.8 & 3289 & 5.8 & 5.9 & 2707 & 15.7 & 4.1 & 354 & 7.6 & 6.8 & 0.93 \\
S\_24 & 6517 & 12.8 & 10.9 & 3230 & 3.9 & 3.3 & 2874 & 22.8 & 9.1 & 412 & 25.2 & 8.5 & 1.02 \\
S\_25 & 6415 & 11.0 & 8.2 & 3223 & 3.7 & 4.0 & 2824 & 20.7 & 6.0 & 368 & 11.9 & 6.6 & 0.99 \\
S\_26 & 6268 & 8.5 & 6.7 & 3331 & 7.1 & 5.4 & 2589 & 10.6 & 3.0 & 347 & 5.5 & 4.6 & 0.88 \\
S\_27 & 6259 & 8.3 & 7.4 & 3307 & 6.4 & 5.2 & 2603 & 11.2 & 3.2 & 347 & 5.5 & 4.4 & 0.89 \\
S\_28 & 5742 & -0.6 & 6.0 & 3305 & 6.3 & 4.9 & 2111 & -9.8 & 3.2 & 325 & -1.2 & 5.2 & 0.74 \\
S\_29 & 6292 & 8.9 & 7.2 & 3325 & 6.9 & 4.4 & 2609 & 11.5 & 3.6 & 356 & 8.2 & 4.7 & 0.89 \\
S\_30 & 6104 & 5.6 & 5.6 & 3483 & 12.0 & 6.2 & 2319 & -0.9 & 4.1 & 301 & -8.5 & 3.0 & 0.75 \\
S\_31 & 6098 & 5.5 & 11.5 & 3360 & 8.1 & 6.3 & 2414 & 3.2 & 7.6 & 322 & -2.1 & 5.4 & 0.81 \\
S\_32 & 6063 & 4.9 & 6.8 & 3482 & 12.0 & 7.2 & 2281 & -2.5 & 4.2 & 299 & -9.1 & 3.4 & 0.74 \\
S\_34 & 6201 & 7.3 & 6.7 & 3376 & 8.6 & 5.5 & 2480 & 6.0 & 3.0 & 345 & 4.9 & 4.8 & 0.84 \\
S\_35 & 5905 & 2.2 & 4.2 & 3244 & 4.3 & 4.4 & 2315 & -1.1 & 3.0 & 345 & 4.9 & 4.5 & 0.82 \\
\hline
\multicolumn{14}{l}{ {\bf Notes:} } \\
\multicolumn{14}{l}{$^{(1)}$ Percentage of the difference between model and observations with respect to the observations: $100\times (N_\mathrm{model} - N_\mathrm{obs.}) / N_\mathrm{obs.}$ }\\
\end{tabular}

%% file: appendixA.tex
\section{Criteria for classifying AGB stars}
\label{app:classification}
 
\subsection{Observational criteria}
SMC stars in B11 and SR16 are classified into red Supergiants (RSG), Carbon-rich AGB (C-AGB), Oxygen-rich AGB (O-AGB), anomalous Oxygen-rich AGB (a-AGB) and extreme AGB (X-AGB). 
C-AGB and O-AGB stars are selected in the \jks\ CMD, following \citet{cioni06a}. All C-AGB and O-AGB stars are brighter than K0 line, defined as:
\begin{equation}
\mathrm{K0} = -0.48\times(\jks) + 13.022 + 0.056\mh
\end{equation}
where the value of the metallicity is $\mh = \log(Z_\mathrm{SMC}/Z_\odot)$. 
The adopted values of metallicity and distance of the SMC are $Z_\mathrm{SMC} = 0.2$  $Z_\odot$ and $d_\mathrm{SMC} = 61$~kpc. Two additional lines separate between C- and O-AGB stars:
\begin{eqnarray}
\mathrm{K1} &=& -13.333\times(\jks) + 25.293 + 1.568\mh 
\label{eq:k1} 
\\
\mathrm{K2} &=& -13.333\times(\jks) + 29.026 + 1.568\mh
\label{eq:k2}
\end{eqnarray}
C-AGB stars have \jks~colors redder than the K2 boundary while O-AGB stars lie between K1 and K2. To minimize the contamination from RGB stars, sources fainter than the TRGB level in both \ks\ and 3.6$\mu$m passbands -- estimated to be at 12.7~mag and 12.6~mag, respectively -- are excluded from the sample.  

The class of the reddest AGB stars is denoted as extreme AGB. The majority of them are probably in the ``superwind'' phase with high mass-loss rates. Since they are obscured at optical wavelengths by thick dusty envelopes, the selection also includes mid-IR photometry. The sources classified as X-AGB stars are brighter than the 3.6$\mu$m TRGB and redder than $\col{J}{[3.6]}>3.1$~mag, or redder than $\col{[3.6]}{[8]} > 0.8$~mag if the $J$-band detection is not available.
Two additional criteria allow to minimize the contamination from YSOs and unresolved background galaxies \citep[see 3.1.2 in][]{boyer11}.

The classification criteria adopted for selecting the RSGs are as follows: (1) they are bluer than the K1 line;  
(2) they have \ks\ brighter than the TRGB and, (3) they are redder than the K$_{\rm{RSG}}$ line which is parallel to the K1 line but shifted by $\Delta (\jks) = -0.2$.
These criteria minimize the contamination from O-rich stars, from RGB stars and foreground/background sources.

\citet{boyer11} identified a new feature in the \cmd{[8]}{J}{[8]} CMD suggesting the presence of a class of stars distinct from C- and O-AGB. They referred to these stars as anomalous Oxygen stars (aO-AGB), as they were originally classified as O-rich by \citet{cioni06a}. These stars were selected from the original O-AGB sample if they are redder than the line ($aO$):
\begin{equation}
[8]= A -11.76\times( \col{J}{[8]} ),
\label{eq:rgbL}
\end{equation}
with $A = 27.95$~mag, and with an 8$\mu$m absolute magnitude fainter than $M_{8}=-8.3$~mag. Stars in the original C-AGB sample fall in this population if they are bluer than the line ($aC$) defined by Eq.~\ref{eq:rgbL} with $A = 31.47$~mag.
In a more recent work, \citet{boyer15} carried out a detailed analysis of chemistry, pulsation properties, stellar parameters and dust production of these stars and concluded that they are low-mass dusty AGB stars at the very end of their evolution, with median current stellar masses of about 0.94~\Msun\ and initial masses $M < 1.25$~\Msun. The spectral classification has been performed for a sample of 273 aO-AGB stars and resulted in 122 C-rich, 100 O-rich, 23 S-type and 28 unknown spectra. Since a high fraction ($\sim45$ per cent) of the aO-AGB stars turned out to be C-rich, \citet{boyer15} refer to aO-stars simply as a-AGB stars. Their results suggest that a-AGB stars should be photometrically selected using the \col{J}{[8]} colour, which is more reliable with respect to \jks\ colour for the most evolved stars. 

\begin{figure} 
\centering
\includegraphics[width=\columnwidth]{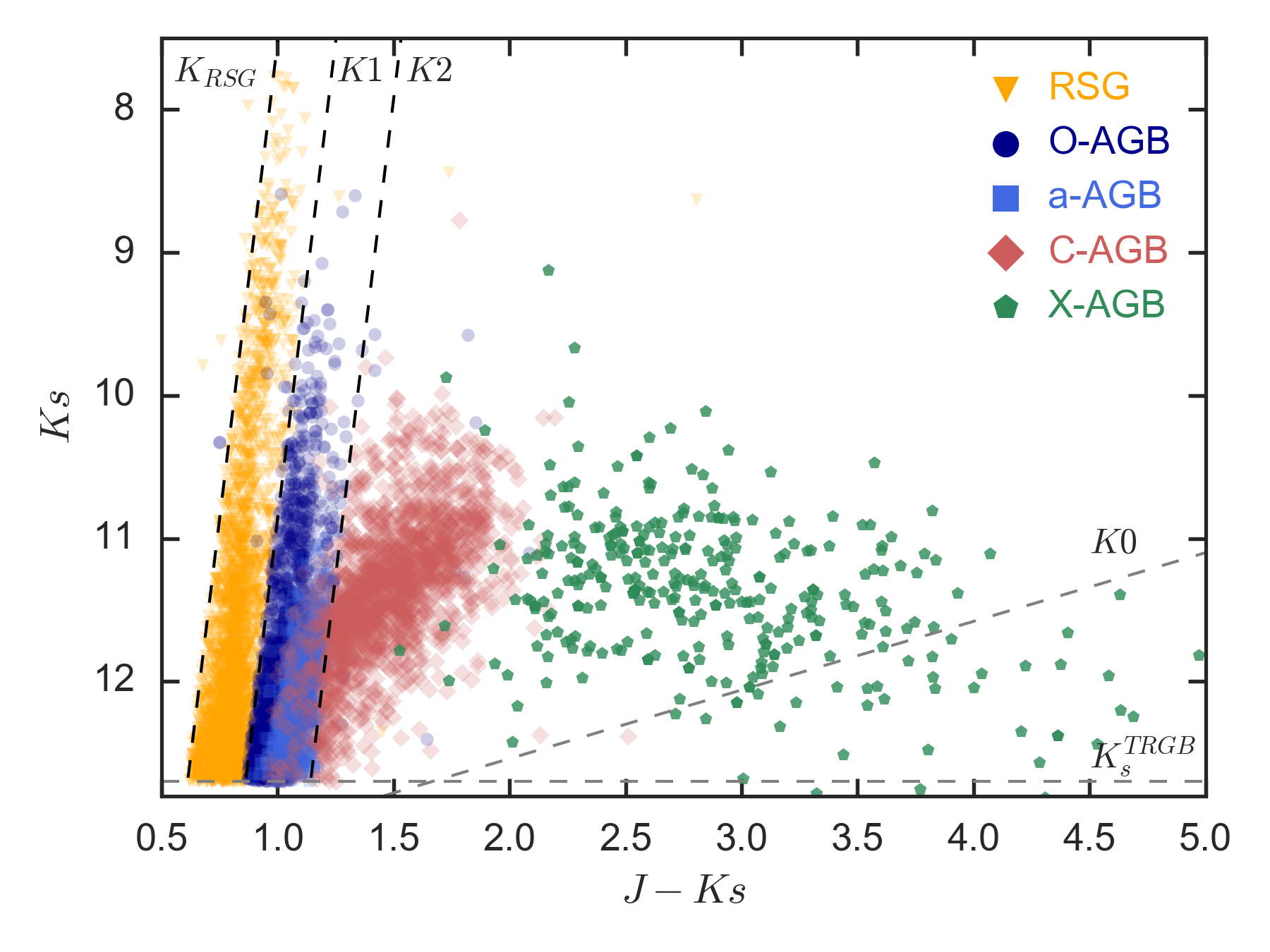}
\includegraphics[width=\columnwidth]{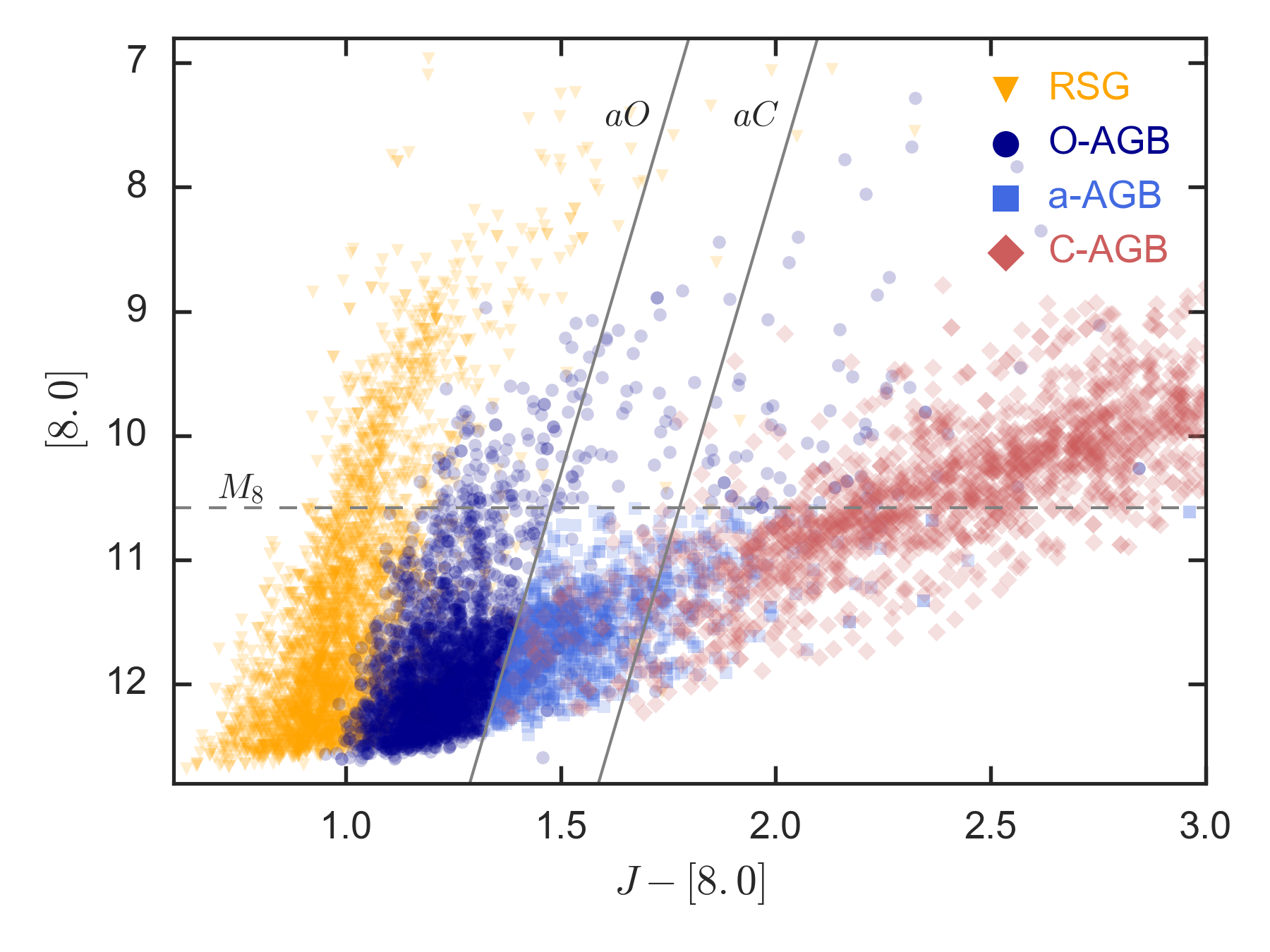}
\caption{ \cmd{\ks}{J}{\ks} (upper panel) and \cmd{[8.0]}{J}{[8.0]} (lower panel) CMDs of the observed sample of evolved stars in the SMC. In the upper panel, grey lines are the photometric criteria adopted to separate RSG from AGB stars and O-rich from C-rich stars (K0, K1 and K2, see text), the horizontal dashed line is the TRGB in the \ks\ band. In the lower panel, the three lines ($aO$, $aC$, $M_{8}$) are the photometric criteria used to select the a-AGB stars as explained in the text. Stars are colour-coded according to their class and plotted with different symbols as shown in the legend.}
\label{fig:SR_KJK_8J8}
\end{figure}

Figure \ref{fig:SR_KJK_8J8} shows the \cmd{\ks}{J}{\ks} CMD of the observed sample of evolved stars in the SMC area used in this work. The grey lines are the photometric criteria adopted to separate RSG from AGB stars and O-rich from C-rich stars. The dashed line is the tip of the RGB in the \ks\ band. Stars are colour-coded according to the evolutionary stage as shown in the legend of the figure. 
 
\subsection{Model selection criteria}
\label{ssec:sel_criteria}

The X-AGB stars can only be selected using photometric criteria, as they do not correspond to a well defined set of intrinsic stellar parameters. Therefore, a star in the synthetic catalogue is identified as X-AGB following the same criteria as B11: 1)  $\col{J}{[3.6]} > 3.1$~mag or $\col{[3.6]}{[8]} > 0.8$ mag if $J > 19.69 $ mag; 2) $[3.6] < [3.6]_\mathrm{TRGB}$.
To identify the RSG sequence we use again the photometric criteria: 1) $Ks < \ks_\mathrm{TRGB}$; 2) $[3.6] < [3.6]_\mathrm{TRGB}$ and 3)  $ K_\mathrm{RSG} <= \ks <= K1$.

A simple comparison between models and observations suggests that the sequence of O-AGB stars is populated by both TP-AGB stars and early-AGB stars, which can be easily distinguished in the synthetic catalogues (looking at their evolutionary stages) but not in the data.
To properly compare the synthetic and observed O-AGB sequences we first select them as in B11: 1) \jks\ redder than $K1$; 2) $\ks < \ks_\mathrm{TRGB}$; 3) $[3.6] < [3.6]_\mathrm{TRGB}$ . We then considered the value of the $\co$ ratio and selected only stars with $\co < 1$.

The C-stars are selected according to the $\co$ ratio, i.e.\ if $\co > 1$. We exclude from the C-AGB sequence the stars fainter than both $\ks_\mathrm{TRGB}$ and $[3.6]_\mathrm{TRGB}$ and the already selected sample of X-AGB stars.  

The observed class of a-AGB stars could be identified in the synthetic catalogues by using the photometric criteria, but this class hosts both C- and O-rich AGB stars that cannot be clearly separated using photometry. The only information we have about their chemical type is the spectral classification performed by \citet{boyer15}. They concluded that in the SMC the ratio between C-rich and O-rich a-AGB stars is $\approx 50$ per cent. When comparing the simulated and observed \ks-band LFs, we take into account the contribution of the a-AGB stars by weighting the  observed C- and O-AGB LFs with the a-AGB LF, according to the result of \citet{boyer15}. Similarly, the total number counts of the C- and O-AGB stars are corrected to include the contribution of the aAGB stars. Therefore, the total number of observed C-AGB and O-AGB changes from 1854 to 2340 and 2623 to 3109 respectively.

%% file: appendixB.tex
\section{The colour shift of giants and supergiants}
\label{app:colorshift}

\begin{figure} 
\centering
\includegraphics[width=\columnwidth]{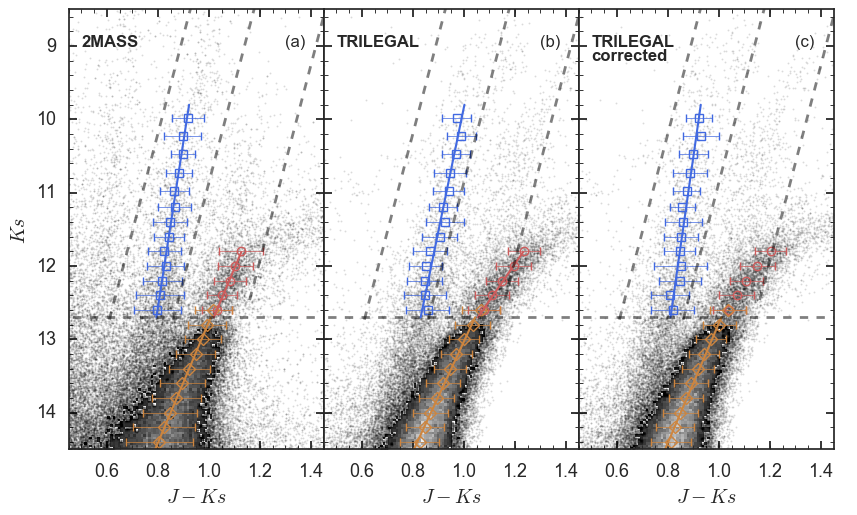}
\caption{Panel (a) \cmd{\ks}{J}{\ks} CMD of the SMC from 2MASS data. Panel (b) synthetic CMD (c) same as in panel (b) but with the correction applied to the synthetic photometry as explained in the text. The 2MASS data and the simulation cover the same SMC areas. The fiducial lines are shown as solid lines 
and the centres of the magnitude bins are marked with different empty symbols for the RGB (diamonds), the AGB (circles) and the RSG (squares) sequences. The error bars represents the $1-\sigma$ standard deviations of the fitted gaussian distributions. The dashed horizontal line marks the \ks-band TRGB, while the diagonal dashed lines correspond to the photometric cuts used by B11 and S11 to separate the RSG sequence from the Milky Way foreground and the sequence of O-rich AGBs from the RSGs. 
}
\label{fig:colorshift_cmd}
\end{figure}

By comparing the observed and simulated \cmd{\ks}{J}{\ks} CMD we found an evident discrepancy in the colour and slope of the RSG and O-rich TP-AGB sequences. The left panels of Fig.~\ref{fig:colorshift_cmd} show the CMD of the SMC from the 2MASS data and the \trilegal\ simulation performed with the TP-AGB set S\_00. 
 By  visual inspection it is immediately evident that the simulated RSG and O-rich AGB sequences are redder than the observed ones.  

\begin{figure} 
\centering
\includegraphics[width=\columnwidth]{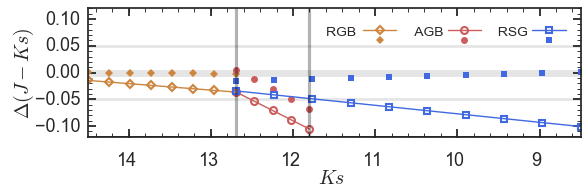}
\caption{\jks\ difference between the observed and simulated sequences as a function of \ks\ for the RGB, AGB and RSG sequences. The differences corresponding to the uncorrected simulation are shown as solid lines with empty markers, while the differences after the correction applied to the photometry are shown as smaller filled symbols. The horizontal lines represent a zero difference in colour and a difference of $\pm$0.05 mag.
}
\label{fig:colorshift_dav}
\end{figure}

To quantify such a shift in colour, we calculate the fiducial lines of both the observed and simulated RSG, O-rich and RGB sequences. The fiducial lines are obtained by fitting a Gaussian distribution in the case of the RGB and a double Gaussian for the RSG and AGB to the distributions of the \jks\ colours in different magnitude bins.  We use a linear fit to infer the value of the shift as a function of the \ks\ magnitude, for the RGB stars ($\ks > 12.7 $), for the AGB stars ($11.8 < \ks < 12.7 $) and for the RSG stars ($\ks < 12.7 $). The linear fits are shown as solid lines.
Figure~\ref{fig:colorshift_dav} shows the \jks\ difference between the observed and simulated sequences as a function of \ks. The colour of the RGB sequence is reproduced within 0.025 mag, while the RSG and the AGB are shifted toward redder colours, with differences spanning from 0.05 to 0.1 mag.

For the purpose of this work, it is important to have the right slope and colours for the different evolutionary sequences because of the photometric selections that have been performed on the data. This is particularly important for the selection of RSG and O-AGB stars for which we rely on the photometric criteria of SR16 and B11.

To reconcile the differences in colours, we correct the synthetic photometry a posteriori by assuming that the \jks\ colour difference is due to an extinction-term correction $dA_{\lambda}$. 
We first derive the  $dA_{J}$ correction from the \jks\ colour difference as a function of the \ks\ magnitude for the RGB, RSG and AGB sequences, by assuming a zero correction for the \ks\ band.
We also calculate the value of $dA_{\lambda}$ as a function of \ks\ magnitude for all the photometric bands bluer than \ks, considering the corresponding extinction coefficients.
The corrections for the $J$-band, $dA_{J}$, are the following:

\begin{eqnarray}
 dA_{J}^\mathrm{RGB} &=&  0.012 \ks  -0.191   ~~  \mathrm{if}  ~ \ks\!>\!12.7 \\
 dA_{J}^\mathrm{RSG} &=&  0.016 \ks  -0.236   ~~  \mathrm{if}  ~  \ks\!<\!12.7  \\
 dA_{J}^\mathrm{AGB} &=&  0.077 \ks  -1.020  ~~  \mathrm{if}   ~ 12.7\!<\!\ks \!<\! 11.8 
\end{eqnarray} 

We find that similar shifts occur also at shorter wavelengths, and that they can be likewise mitigated by applying such corrections.  

In Figure~\ref{fig:colorshift_cmd} we show the simulated CMD after correcting the sequences and the resulting \jks\ colour differences. The differences for the RGB and RSG sequences are reduced to below 0.01 mag, while a slightly larger difference (up to 0.05 mag) is still present in the AGB sequence.

Such residual difference is due to the fact that in the synthetic catalogues, we apply the correction derived for the RSG sequence to the stars brighter than $\ks = 12.7$ mag and the correction derived for the RGB sequence to the stars fainter than $\ks = 12.7$ mag. 
This choice is motivated by the fact that RSG and RGB models are computed with the \parsec\ code and they are not the subject of the present calibration.

Given their wavelength-dependent behaviour, it is unlikely that these shifts are due to some abundance or opacity effect. They could well be caused by small temperature offsets. Independently of their possible cause, we emphasize that these shifts do not significantly affect the results of our calibration.